\documentclass[a4paper,fleqn,usenatbib]{mnras}

\usepackage[T1]{fontenc}
\usepackage{ae,aecompl}

\usepackage{graphicx}	
\usepackage[caption = false]{subfig}
\usepackage{amsmath}	
\usepackage{amssymb}	
\usepackage{pdflscape}
\usepackage{deluxetable}

\title[Mass modelling of the superthin galaxy, FGC1540]{Mass modelling of a superthin galaxy, FGC1540}

\author[Kurapati et al.]{
Sushma Kurapati $^{1}$,\thanks{E-mail: sushma@ncra.tifr.res.in}
Arunima Banerjee $^{2}$, 
Jayaram N. Chengalur $^{1}$,
Dmitry Makarov $^{3}$,
\newauthor
Svyatoslav Borisov $^{4,5}$,
Anton Afanasiev $^{4,5}$,
Aleksandra Antipova $^{3}$ 
 \\
$^{1}$ National Centre for Radio Astrophysics, Tata Institute of Fundamental Research, PO Box 3, Pune 411007, India\\
$^{2}$ Indian Institute of Science Education and Research, Rami Reddy Nagar, Karakambadi Road,
Mangalam (P.O.), Tirupati 517507, India\\
$^{3}$ Special Astrophysical Observatory of RAS, Nizhnij Arkhyz 369167, Karachai-Cherkessian Republic, Russia\\
$^4$ Sternberg Astronomical Institute, Moscow M.V. Lomonosov State University, Universitetsky pr., 13, Moscow 119234, Russia\\
$^{5}$ Department of Physics, Moscow M.V. Lomonosov State University, 1, Leninskie Gory, Moscow 119991, Russia
}

\date{Accepted XXX. Received YYY; in original form ZZZ}

\pubyear{2016}

\begin{document}
\label{firstpage}
\pagerange{\pageref{firstpage}--\pageref{lastpage}}
\maketitle

\begin{abstract}

We present high resolution H{\sc i} 21cm Giant Meterwave Radio Telescope (GMRT) observations of the superthin galaxy FGC1540 with a spatial resolution of 10$''$  $\times$ 8$''$ and a spectral resolution of 1.73 kms$^{-1}$ and an rms noise of 0.9 mJy per beam. We obtain its rotation curve as well as deprojected radial H{\sc i} surface density profile by fitting a 3-dimensional tilted ring model directly to the H{\sc i} data cubes by using the publicly-available software, Fully Automated Tirrific (FAT). We also present the rotation curve of FGC1540 derived from its optical spectroscopy study using the 6-m BTA telescope of the Special Astrophysical Observatory of the Russian Academy of Sciences. We use the rotation curve, the H{\sc i} surface density profile together with Spitzer 3.6 $\mu$m and the SDSS $i$--band data to construct the mass models for FGC1540. We find that both the Pseudo-isothermal (PIS), as well as Navarro-Frenk-White (NFW) dark matter (DM) halos, fit the observed rotation curve equally well. The PIS model indicates a compact dark matter halo ($R_{\rm C}/R_{\rm D}$ < 2), with the best-fitting core radius ($R_{\rm C}$) approximately half the exponential stellar disc scale length ($R_{\rm D}$), which is in agreement with the mass models of superthin galaxies studied earlier in the literature. Since the vertical thickness of the galactic stellar disc is determined by a balance between the net gravitational field and the velocity dispersion in the vertical direction, the compact dark matter halo may be primarily responsible in regulating the superthin vertical structure of the stellar disc in FGC1540 as was found in case of the superthin galaxy UGC7321.

\end{abstract}




\section{Introduction}

Superthin galaxies are galaxies which when viewed edge-on exhibit highly flattened stellar discs with large disc axial ratios ($a$/$b$ $\gtrsim$ 10) \citep{goad81,karachentsev93, karachentsev99}. These galaxies are late-type and gas-rich low surface brightness (LSB) galaxies with little or no bulge component \citep[see][for a review]{kautsch09}. These features suggest that these galaxies are some of the least-evolved galaxies and are, therefore, the ideal test-beds to study the early stages of quiescent disk galaxy evolution, and to constrain models of galaxy formation $\& $ evolution. 

Superthin galaxies are mostly found in the group and isolated environments \citep{kautschetal09}. It is now well-known that galaxy mergers are common in groups, 
and should trigger the formation of central spheroids or bulges in galactic discs; therefore theoretical models predict that not many simple discs should survive the cosmological evolution of galaxies \citep{naab03, donghia06, khochfar09}. Despite this, pure discs do exist and, in fact, constitute a significant fraction (~15-18 percent) of the observed disc galaxy population \citep{kautscha09, kautsch06}, a fact not well-understood within the framework of the current cosmological paradigm of hierarchical galaxy formation  \citep{kautsch09}. The survival of a superthin vertical structure in some of these pure stellar discs despite the expected heating by stellar bars and spiral arms and other mechanisms relevant for galactic secular evolution is puzzling.


A viable explanation for the existence of superthin discs could be that their discs are exceptionally stable. Theoretical models predict that superthin galaxies must be hosted in massive dark matter haloes to stabilize their discs against perturbations \citep{zasov91, gerritsen99} and, in fact, a recent study revealed a correlation between the thickness of stellar discs and the relative mass of the dark matter halo. \citep{zasov02, kregel05, mosenkov10}.   \citet{banerjee10} found that UGC7321, a prototypical superthin galaxy,  \citep[see e.g.,][]{matthews99, matthews00, matthews01, uson03} has a dense and compact dark matter halo with $R_{\rm C}$/ $R_{\rm D}$ < 2, where $R_{\rm C}$ is the core radius of the Pseudo Isothermal (PIS) dark matter halo and $R_{\rm D}$ is the exponential stellar disc scale length \citep[See also][]{obrien10}. This is in contrast to the high surface brightness (HSB) galaxies, for which the core radius is 3-4 times the exponential stellar disc scale length \citep{gentile04, narayan05, banerjee08}.  \citet{banerjee13} dynamically modelled the vertical thickness of the stellar disc of  UGC7321. The vertical thickness of the stellar disc is determined by the balance between the vertical gravitational force and the pressure. They found that while the vertical force field due to the gas self-gravity does constrain the stellar disc thickness, it is the compact dark matter halo which strongly regulates the mean distribution of stars in the vertical direction in low-luminosity bulgeless galaxies, and is primarily responsible for the stellar disc to be superthin. In addition, \citet{ghosh14} also found that gravitational instabilities in UGC7321 are suppressed by its dark matter halo. In fact, using linear disc stability analysis studies, \citet{garg17} indicated that the dark matter halo is crucial for stabilizing the disc in low surface brightness galaxies in general. Thus, the compact nature of the dark matter halo predominantly determines the disc vertical structure, which may have important implications for the early evolution of galaxies.
 

Besides UGC7321, only two other superthin galaxies have been studied in detail: IC5249 \citep{vander01, banerjee17} and IC2233 \citep{banerjee17}. 
In this paper, we study FGC1540, a proto-typical superthin galaxy in detail.
We present the results of H{\sc i} 21cm radio-synthesis observations from the Giant Meterwave Radio Telescope (GMRT) of FGC1540; we also model its rotation curve and de-projected H{\sc i} surface density and finally construct the mass models using photometrically calibrated $g$, $r$, and $i$ band images from the SDSS Data Access Server and Spitzer 3.6 $\mu$m data in conjunction with the H{\sc i} data.

The rest of the paper is organized as follows: In \S \ref{obs}, we describe the H{\sc i} observations and data analysis, in \S \ref{photometry} the optical and IR photometry, 
in \S \ref{kinematics} H{\sc i} surface density and rotation curve, in \S \ref{models} the mass models followed by results and discussion in \S \ref{results}. Finally, we summarize the main results in \S \ref{summary}.

\section{H{\sc i} Observations $\&$ Data analysis}
\label{obs}

The galaxy FGC1540 was observed with the GMRT over multiple sessions. The observations were carried out on 25, 26 April, 29 August, 09 September, 06 and 08 April in 2013 and a total of 32 hours were spent on the source. The parameters of the GMRT observations are summarized in Table \ref{table_gmrt}. The data was processed using the standard tasks in AIPS package. Bad visibility points for each scan were flagged out to begin with, following which the data was calibrated. 
The task CVEL was performed on all the calibrated data sets in order to correct for the Doppler shift due to earth's motion and then the data sets on all runs were combined using the task DBCON . The task UVSUB was then used to subtract the continuum emission by interpolating from the line free channels in the uv plane. After the continuum subtraction, the data cubes were made at various resolutions using the task IMAGR. The integrated H{\sc i} flux was measured from the data cube by using the Source Finding Application (SoFiA; \citet{serra15}).

\begin{table}
\small
\caption{ Parameters of H{\sc i} observations}
\label{table_gmrt}
\begin{tabular}{ p{4.0cm} p{3.5cm}  }
\\
\hline
 Parameters &  Value  \\ 
\hline
  RA(2000) &  13$^{h}$02$^{m}$08.1$^{s}$   \\  
  Declination(2000) & +58$^{\circ}$42$'$05$''$ \\
  Date of observations & 25 and 26 Apr, 29 Aug, 09 Sep, 06 and 08 Nov, 2013 \\  
  Flux calibrator & 3C147, 3C286  \\
  Phase calibrator & 1400+621 \\
  Central velocity(heliocentric) & 670 km s$^{-1}$  \\
  Time on source & 32h  \\
  Total bandwidth & 4.0 MHz  \\
  Number of channels & 512  \\
  Velocity resolution & 1.73 km s$^{-1}$  \\
  FWHM of synthesized beam &  $10''\times 8'' $  \\
  RMS noise per channel &  0.9 mJy \\
 
\hline 
  
\end{tabular}

\end{table}

\section{Optical $\&$ IR Photometry}
\label{photometry}

\subsection{SDSS Data}
\label{sdss}

\begin{table}
\caption{Measured photometric parameters of the galaxy}
\label{table_sdss}
\begin{tabular}{ccc}
\hline
\hline
SDSS Band & $m$ & $\mu_{\circ}$  \\ 
          & (mag) & (mag arcsec$^{-2}$) \\
\hline
 $g$ & $13.86\pm0.03$ & $21.42\pm0.10$ \\
 $r$ & $13.48\pm0.03$ & $20.91\pm0.10$ \\
 $i$ & $13.29\pm0.03$ & $20.71\pm0.08$ \\
\hline
\hline 
\multicolumn{3}{p{0.45\textwidth}}{
Notes: The total magnitude, $m$, 
and the central surface brightness, $\mu_{\circ}$,
are not corrected for the foreground extinction.
}
\end{tabular}
\end{table}

The photometrically calibrated Sloan Digital Sky Survey (SDSS) frames for the galaxy FGC 1540 in the $g$, $r$, and $i$ bands were taken from the SDSS Data Access Server \citep{ahn12}. Background subtraction was performed in all the $g$, $r$, and $i$ frames. For this purpose, all the stars and other objects (including FGC 1540) and their surroundings were masked. The regions for masking were chosen based on 3$\sigma$ deviation from the median value of the frame. Regions around the galaxy were masked manually. For every frame, the background was approximated with a second order two-dimensional polynomial and was subtracted.

Elliptical apertures are not the ideal fits to the galaxy isophotes as the galaxy FGC 1540 is a nearly edge-on galaxy. Hence, the total magnitude was estimated by integrating the light in a rectangular box, the size of the box being chosen in such a way that the growth curves reach its asymptotic value near the borders of the box. The measured magnitudes and the central surface brightness in $g$, $r$ and $i$--bands are given in Table \ref{table_sdss}.  This galaxy region was also used for fitting the vertical and radial photometric profiles. 

The data were found to be consistent with a model comprising a thick and a thin disc, each with a $\rm{sech}$ profile in the vertical and an $\rm{exponential}$ profile in the radial direction. Besides, the disc thickness was found not to vary significantly as a function of galactocentric radius $R$. Integrated along the line-of-sight to fit the observed intensity profiles of edge-on systems like FGC1540, the above model results in an intensity profile given by $I = I_{0,0} \ R/R_D  \ K_{1}(R/R_D) \ \rm{sech}$(${z/z_0}$), where $K_{1}$ is the modified Bessel function of the second kind, $R$ the galactocentric radius, $R_D$ the exponential disc scale length, $z_0$ the vertical scale height and $I_{0,0}$ is the intensity at $R$=0 and $z$=0 \citep{vander81}. The commonly used sech$^{2}$($z/z_0$) model gives a worse fit than the sech($z/z_0$) in the vertical direction. The sech($z/z_0$) model was proposed by \citet{vander88} as a model intermediate between the isothermal and exponential models. The star-forming regions from the middle plane were excluded to minimize effects due to dust extinction; only regions located 12$\arcsec$ above and below the middle plane were used to fit the radial profile. The Galactic (Milky Way) dust extinction \citep{schlafly11} was taken into account for the derivation of model-dependent photometric parameters. In general, superthin galaxies suffer very little from dust extinction in spite of their edge-on geometry \citep[see e.g.][]{matthews99}. The dust line in edge-on galaxies becomes visible if the rotation velocity of a galaxy exceeds 120 km s$^{-1}$ \citep{dalcanton04} and the maximum rotation velocity of FGC 1540 is 90 km s$^{-1}$. Hence, the internal dust extinction in FGC1540 is expected to be low and was not corrected for.

The best-fitting parameters of our model of the stellar disc are given in Table \ref{table_opt_par}. The mass-to-light ratio (${\gamma}^{*}$) was calculated using the ($g$-$i$) color of the discs following \citet{zibetti09} and was found to be $\sim$ 0.54 for the galaxy FGC1540. Our results show that the thin disc has a scale length of 1.29 $\pm$ 0.01 kpc, a scale height of 0.185 $\pm$ 0.002 kpc and a total mass of $\sim $ (1.1 $\pm$ 0.07) $\times$ 10$^{8}$ M$_{\odot}$. The thick disc has a scale length of 1.29 $\pm$ 0.01 kpc, a scale height of 0.675 $\pm$ 0.035 kpc and a total mass of $\sim $ (0.87 $\pm$ 0.15) $\times$ 10$^{8}$ M$_{\odot}$. The total stellar mass of the galaxy is $\sim $ 1.97  $\times$ 10$^{8}$ M$_{\odot}$. The absolute B-band magnitude was calculated from the magnitudes at $g$ and $r$--band using the Lupton's transformation assuming a distance of $\sim$ 11 Mpc. The absolute B-band magnitude ($M_{\rm B}$), blue luminosity ($L_{\rm B}$), and the ratio of H{\sc i} mass to the blue luminosity ($M_{\rm HI}$/$L_{\rm B}$) are shown in Table \ref{table_par}. The errors on absolute magnitude are dominated by distance uncertainty ($\sim$ 10 $\%$). 

\subsection{Spitzer IRAC 3.6 $\mu$m}
\label{spitzer}

 One of the main uncertainties in constructing mass models is the conversion from stellar luminosity profiles to stellar mass profiles, using a mass-to-light ratio ($\gamma_{\ast}$) as determined from stellar population synthesis models \citep[See e.g.][]{bell01}. 
The uncertainties in the derivation of $\gamma_{\ast}$ are much less in the mid-infrared compared to the optical case as the latter is dominated by young stellar population, recent star formation events and the effects of dust. Therefore, the Spitzer IRAC 3.6 $\mu$m images serve as a better tracer of the underlying stellar population.


The 2-dimensional structural decomposition of the Spitzer 3.6 $\mu$m intensity profiles was already available in the literature \citep{salo15}. 
FGC1540 was found to be consistent with a model consisting of a superposition of a thin disc and a thick disc, with each of the discs having an exponential radial profile and a sech$^{2}$ vertical profile. 
The surface brightness profiles were converted to mass density profiles assuming a mass-to-light ratio  $\gamma_{\ast}$ of 
0.35, the average $\gamma_{\ast}$ value as calculated for a sample of superthin galaxies \citep{vander01, banerjee17}. We find that the thin disc has a central 
density of 8.16 M$_{\odot}$ pc$^{-2}$, has a scale length of 0.54 kpc and a scale height of 0.152 kpc; the thick disc has a central density of 3.24 
M$_{\odot}$ pc$^{-2}$, has a scale length of 1.85 kpc and a scale height of 0.43 kpc. The total stellar mass obtained from the Spitzer 3.6 $\mu$m intensity profiles is $\sim$ 0.95 $\times$ 10$^{8}$ M$_{\odot}$, which is $\sim$ 2 times lower as compared to the estimate from the SDSS $i$--band observations. The structural parameters derived using the SDSS $i$--band and the Spitzer 3.6 $\mu$m data are summarized in Table~\ref{table_opt_par}.

\begin{table}
\small
\caption{Surface brightness decompositions at SDSS $i$--band and Spitzer 3.6 $\mu$m band}
\label{table_opt_par}
\begin{tabular}{ p{1.7cm} p{1.3 cm} p{0.6 cm} p{0.6cm} p{0.6cm} p{1.0cm}  }
\\
\hline
\hline
 Band & disc & $\mu_{\circ}$  & $R_{\rm D}$ & $h_{\rm z}$ & $M_{\ast}$ \\ 
  \hline
 3.6 $\mu$m & thin disc & 21.39 & 0.54 & 0.152 & 0.15 \\
   & thick disc& 22.23 & 1.85 & 0.43 & 0.80\\
\hline
 
 SDSS $i$--band & thin disc  & 20.60 & 1.29  & 0.185 & 1.10\\
  & thick disc & 21.67 & 1.29  & 0.675 & 0.87\\

\hline
\hline 
\multicolumn{6}{l}{Notes: $\mu_{\circ}$ is face-on surface brightness in mag arcsec$^{-2}$, $R_{\rm D}$ }\\
\multicolumn{6}{l}{ is the scale length in kpc, $h_{\rm z}$ is the scale height in kpc and}\\
\multicolumn{6}{l}{ $M_{\ast}$ is the stellar mass in units of $\times$ 10$^{8}$ M$_{\odot}$.}
\end{tabular}

\end{table}

\begin{figure}
\centering
\includegraphics[width=1.0\linewidth]{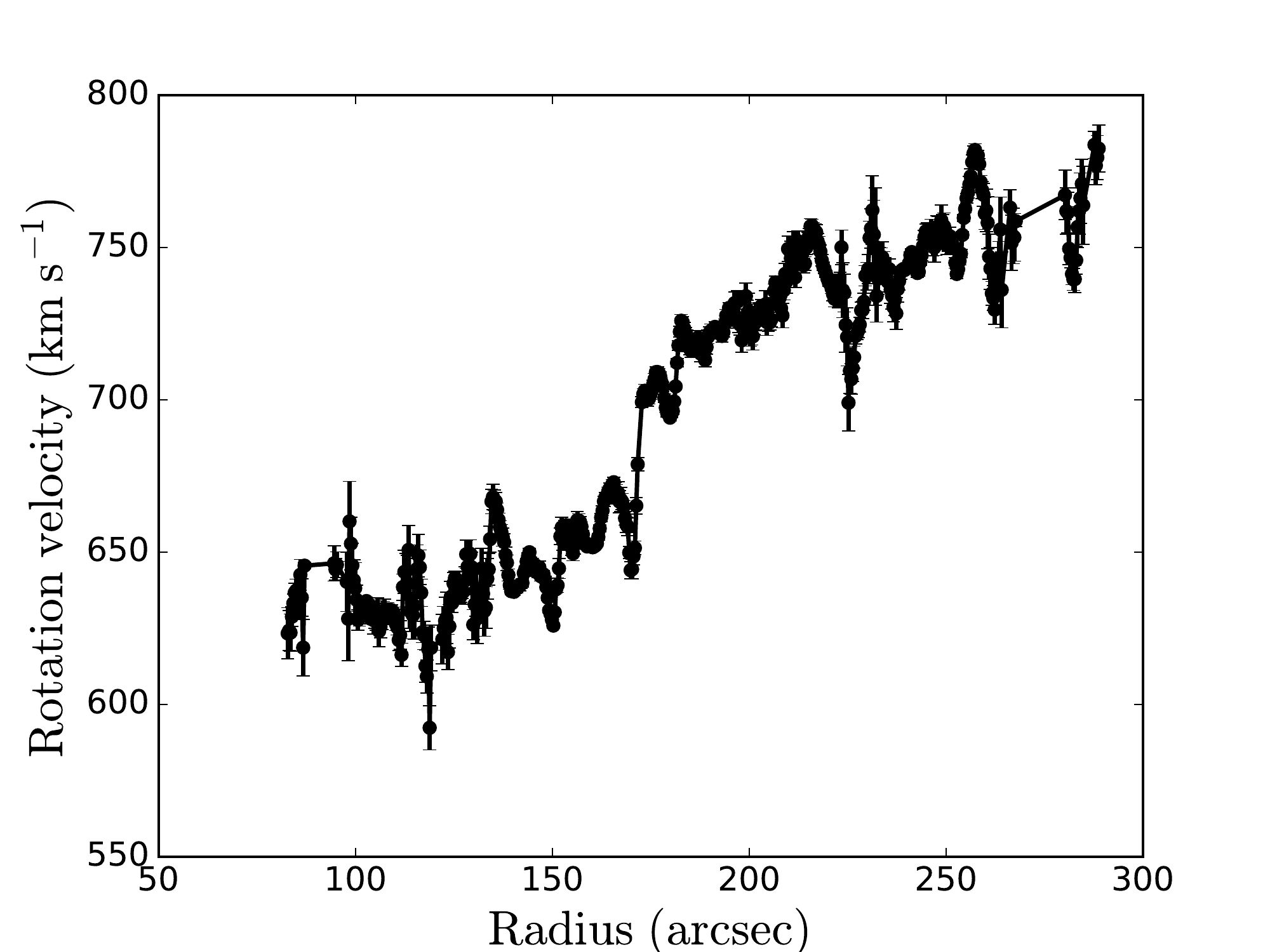}
\includegraphics[width=1.0\linewidth]{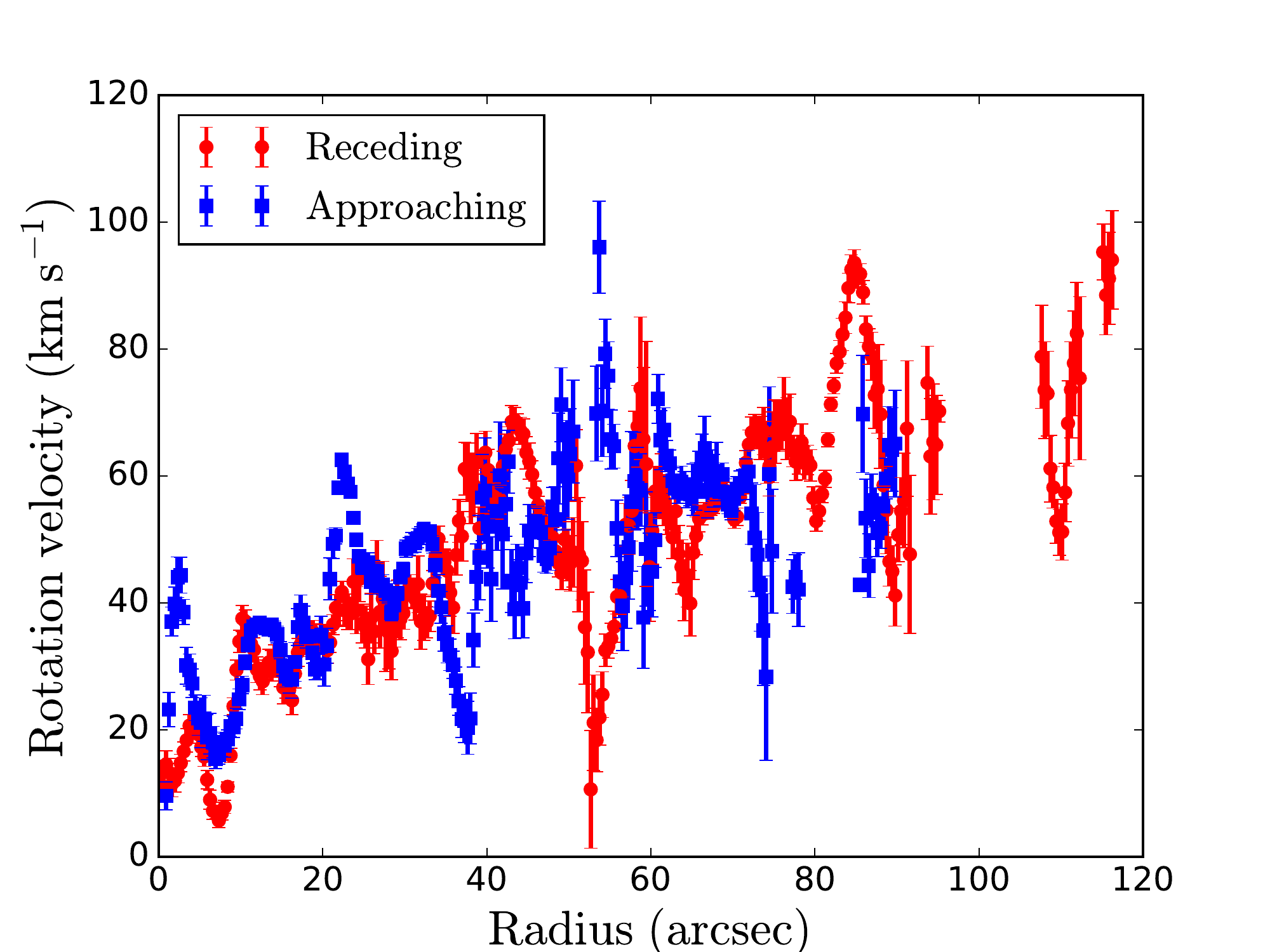}
\caption{The kinematics of the galaxy FGC 1540 using the SAO RAS 6-m telescope (a) The line of sight velocity(km s$^{-1}$) along the galaxy major axis versus the position along the slit (arcsec). (b) The rotation curve using the approaching  half is shown by filled blue squares and the rotation curve using the receding half is shown by open red circles.}
\label{fig:optrot}
\end{figure}%

\subsection{Optical spectroscopy}

The spectroscopic observations of FGC 1540 were obtained on 1st August 2017 with the 6-m BTA telescope at the Special Astrophysical Observatory of the Russian Academy of Sciences with the multi-mode focal reducer  SCORPIO using the VPHG1800R grism along with the CCD detector EEV CCD42-40. A 1$''$-wide slit was oriented along the galaxy's major axis. This gives the reciprocal dispersion about 0.52 {\AA}/pix in range from 6100--7100 {\AA} and FWHM of 2.5 {\AA}. The total exposure time was 1200 sec and the seeing was 1.5 arcsec. The image scale is 0.357 arcsec per pixel along a slit. The standard reduction of the spectra including bias subtraction, wavelength calibration, and redshifts measurements was performed using ESO MIDAS system for image processing with the context LONG.

The typical error of the velocity measurement is 3.5 km~s$^{-1}$. Fig \ref{fig:optrot}~(a) shows the rotation curve along the galaxy major axis.  Fig \ref{fig:optrot}~(b) shows the rotation curve derived for the approaching half (shown by filled blue squares) and for the receding half (shown by open red circles). The dynamical center of the galaxy is found by minimizing the scatter between receding and approaching wings of the rotation curve. The systemic velocity was found to be 689 km s$^{-1}$. The rotation curve shows drops and rises with amplitude up to 40--50 km s$^{-1}$. We interpret this as follows. The optical spectrum is dominated by emission from star forming regions, which are likely to be randomly distributed along the line of sight. Since for edge on galaxies, the mapping between the projected and true galactocentric distance is degenerate, this random distribution could lead to the observed features in the derived rotation curve. We do not expect the rotation curve to be affected by dust, both because, as discussed above, superthin galaxies generally have little dust, and also because the weak 22 $\mu$m emission from FGC1540 indicates that the amount of dust (with temperature of 70--100 K) is quite small in FGC1540.

\begin{table}
\small
\caption{ Properties of FGC1540}
\label{table_par}
\begin{tabular}{ p{4.0cm} p{3.5cm}  }
\\
\hline
 Properties &  Value  \\ 
\hline
  Hubble type & Sd \\  
  Distance$^{a}$ & 11 Mpc  \\
  Systemic velocity $^{b}$ & 670 km s$^{-1}$  \\
  $M_{\rm B}$ $^{c}$& -16.04$\pm$0.19 mag \\
  H{\sc i} Integrated flux & 31.5$\pm$3.1 Jy km s$^{-1} $ \\
  Maximum velocity & 90 km s$^{-1} $\\
  Position angle$^{b}$ & 31$^{\circ}$ $\pm$ 1$^{\circ}$ \\
  Inclination$^{b}$ & 87$^{\circ}$ $\pm$ 1.5$^{\circ}$ \\
  $a/b$ $^{d}$ & 7.47 \\
  $M_{\rm HI}$ & (9 $\pm$ 2) $\times$ 10$^{8}$ M$_{\odot}$ \\
  $L_{\rm B}$ & 2.23 $\times$ 10$^{8}$ L$_{\odot}$ \\
  $M_{\rm HI}$/$L_{\rm B}$  & 4.1 \\
  
\hline 
\multicolumn{2}{l}{  Notes: $^{a}$ Distance was calculated from the Hubble law.   }\\
\multicolumn{2}{l}{ $^{b}$ Systemic velocity, Position angle, and inclination angle were  }\\
\multicolumn{2}{l}{estimated from H{\sc i} kinematics using FAT software.}\\ 
\multicolumn{2}{l}{ $^{c}$ Absolute magnitudes at $B$-band was calculated from the }\\ 
\multicolumn{2}{l}{brightness in $g$ $\&$ $r$ bands using Lupton transformation. }\\
\multicolumn{2}{l}{ $^{d}$ Axial ratio measured at $g$-band.}\\ 
\end{tabular}

\end{table}

\section{H{\sc i} distribution $\&$ kinematics}
\label{kinematics}
\begin{figure}
\centering
\includegraphics[width=1.0\linewidth]{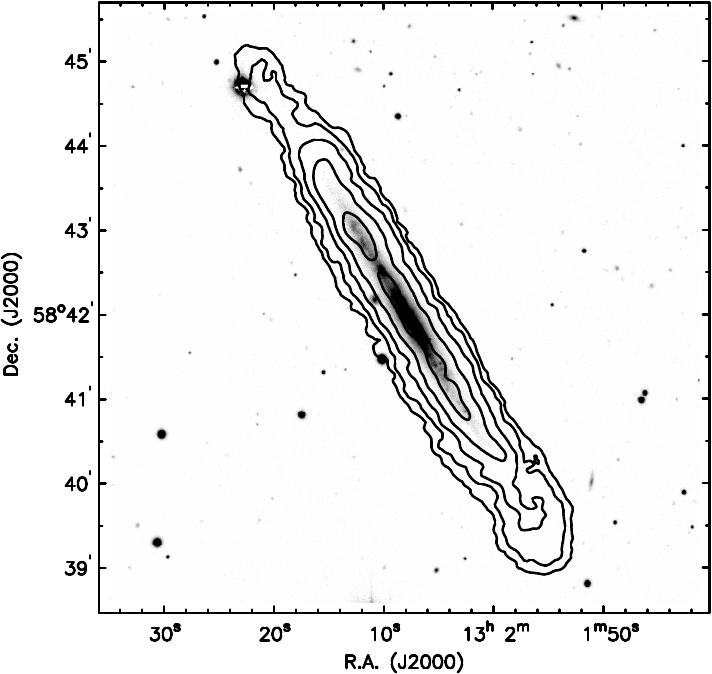}
\includegraphics[width=1.0\linewidth]{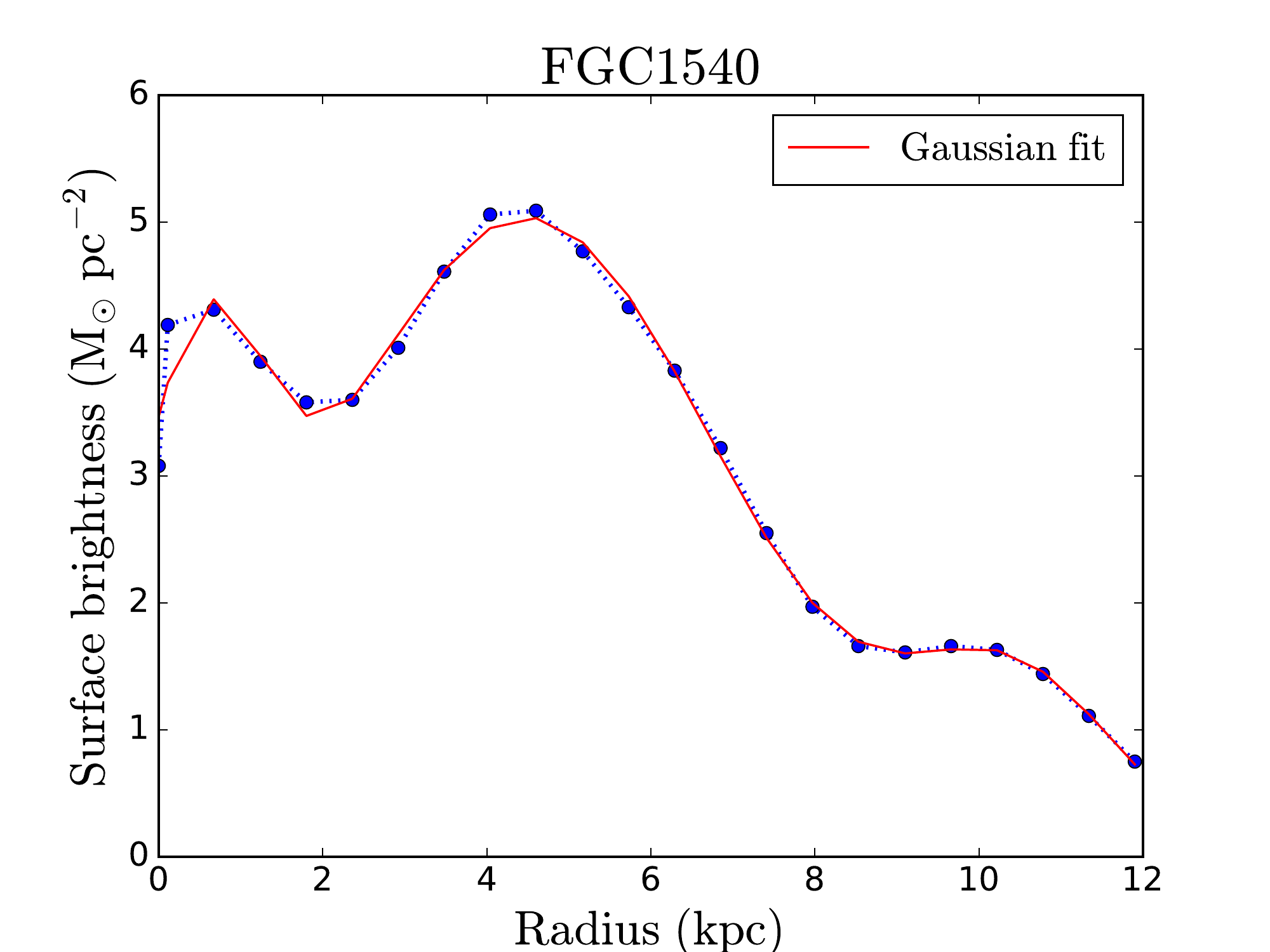}
\caption{(a) H{\sc i} distribution of the galaxy FGC1540 at a resolution of 10$''$ $\times$ 8$''$, overlayed on $g$-band optical image. The contour levels are (3, 6, 12, 24, 48, 96) $\times$ 10$^{20}$  atoms per cm$^{-2}$. The outermost contours correspond to $\sim$ 3$\sigma$. (b) The de-projected H{\sc i} radial surface density profile of the galaxy FGC1540. The red solid line shows the fit with three Gaussians.}
\label{fig:mom0}
\end{figure}

Fig \ref{fig:mom0}~(a) shows the H{\sc i} contours overlayed on the g-band optical image of FGC1540. The H{\sc i} morphology of FGC1540 shows a relatively symmetric distribution extending beyond the stellar disc. We calculate the H{\sc i} mass using the standard formula, $M_{\rm HI} $= 2.36$\times$ 10$^{5}$ $D^{2} \int S$ d$v~ M_{\odot}$, where $D$ is the distance in Mpc, $S$ is the flux density in Jy and d$v$ is in km s$^{-1}$. We find an integrated flux of $\sim$ 31.5 Jy km/s, which corresponds to a H{\sc i} mass of $\sim$ 9 $\times$ 10$^{8}$ M$_{\odot}$ by assuming a distance of 11 Mpc. The estimated H{\sc i} integrated flux density and the H{\sc i} mass for the galaxy FGC1540 are given in Table \ref{table_par}. The uncertainty on integrated flux density are dominated by the calibration uncertainties, which for the GMRT are typically $\sim$ 10 $\%$. The error on the H{\sc i}  mass includes the errors on the distance measurements ($\sim$ 10 $\%$) as well as the error on flux density.

The kinematics of disc galaxies are traditionally modelled using the `tilted ring model' \citep{rogstad74}. Derivation of rotation curves and the de-projected H{\sc i} surface densities for edge on galaxies is not possible by the usual 2D tilted ring fits which work on 2D velocity fields \citep[e.g. ROTCUR in the GIPSY package,][]{vanderhulst92}. These 2D tilted ring models work well for galaxies with inclinations lower than 70$^{\circ}$ \citep[e.g.][]{begeman89,oh18}. 
 There are now packages (e.g., TIRIFIC, \citet{jozsa07}; 3d-BAROLO \citet{diteodoro15})which can derive the rotation curve by directly fitting to the 3-D data cube. Since 3D fitting techniques are the most suitable for extracting kinematical information from edge-on galaxies \citep[e.g.][]{kamphuis13}, we use Fully Automated TIRIFIC (FAT) software \citep{kamphuis15} to derive the de-projected H{\sc i} surface brightness profile and the rotation curve of FGC1540.

\subsection{FAT}
We derive the rotation curve of FGC1540 by fitting a tilted ring model to the  3-D data cube using the FAT software. FAT is a wrapper around TIRIFIC \citep{jozsa07}. FAT also uses SOFIA \citep{serra15} to obtain estimates of the initial parameters for the rings. The FAT pipeline fits a 3D tilted ring model directly to the 3-D data cube and determines the rotation curve. FAT errors are empirical errors that are determined by taking the maximum of the difference between the unsmoothed and the smoothed profile, the variation of the smoothed profiles in the Monte Carlo fitting process and a minimum default value. In case of the rotation curve, this minimum value is set as the maximum of 5 km s$^{-1}$ or 0.5$\times$ d$v$/ $\sin (i)$, where d$v$ is the velocity resolution and $i$ the inclination (see \citet{kamphuis15} for more details on error determination). 

 The de-projected H{\sc i} radial surface brightness profile for the galaxy FGC1540 is shown in Fig. \ref{fig:mom0}~(b) by the blue dotted line. We fit a triple Gaussian profile to the radial surface density distribution which is shown by the red solid line. The H{\sc i} parameters for the galaxy FGC1540 are listed in Table \ref{table_par}.

\subsection{Correction for pressure support}
The observed rotation velocities under-estimate the dynamical mass when the support due to the radial gas pressure gradient is not insignificant. Hence,
in order to construct the mass models, the observed rotation velocities have to be corrected for the pressure support \citep[e.g.][]{meurer96, begum04}. 
This correction is given by:
\begin{equation}
v_{c}^{2} = v_{o}^{2} - r \times \sigma ^{2} \bigg[\frac{d}{dr}(ln \  \Sigma_{HI}) \ + \frac{d}{dr}(ln \  \sigma^{2}) \ - \frac{d}{dr}(ln \  2h_{z}) \bigg]
\end{equation}
where, $v_{c}$ is the corrected velocity, $v_{o}$ is the observed rotation velocity, $\Sigma_{HI}$ is the H{\sc i} surface density, $\sigma$ is the velocity dispersion and $h_{z}$ is the scale height of the disc. We assume that the scale height does not vary with radius and that the velocity dispersion is constant across the galaxy, which gives:
\begin{equation}
\label{correction}
v_{c}^{2} = v_{o}^{2} - r \times \sigma^{2} \bigg[\frac{d}{dr}(ln \  \Sigma_{HI}) \bigg]
\end{equation} 

\begin{figure}
\centering
\includegraphics[width=1.0\linewidth]{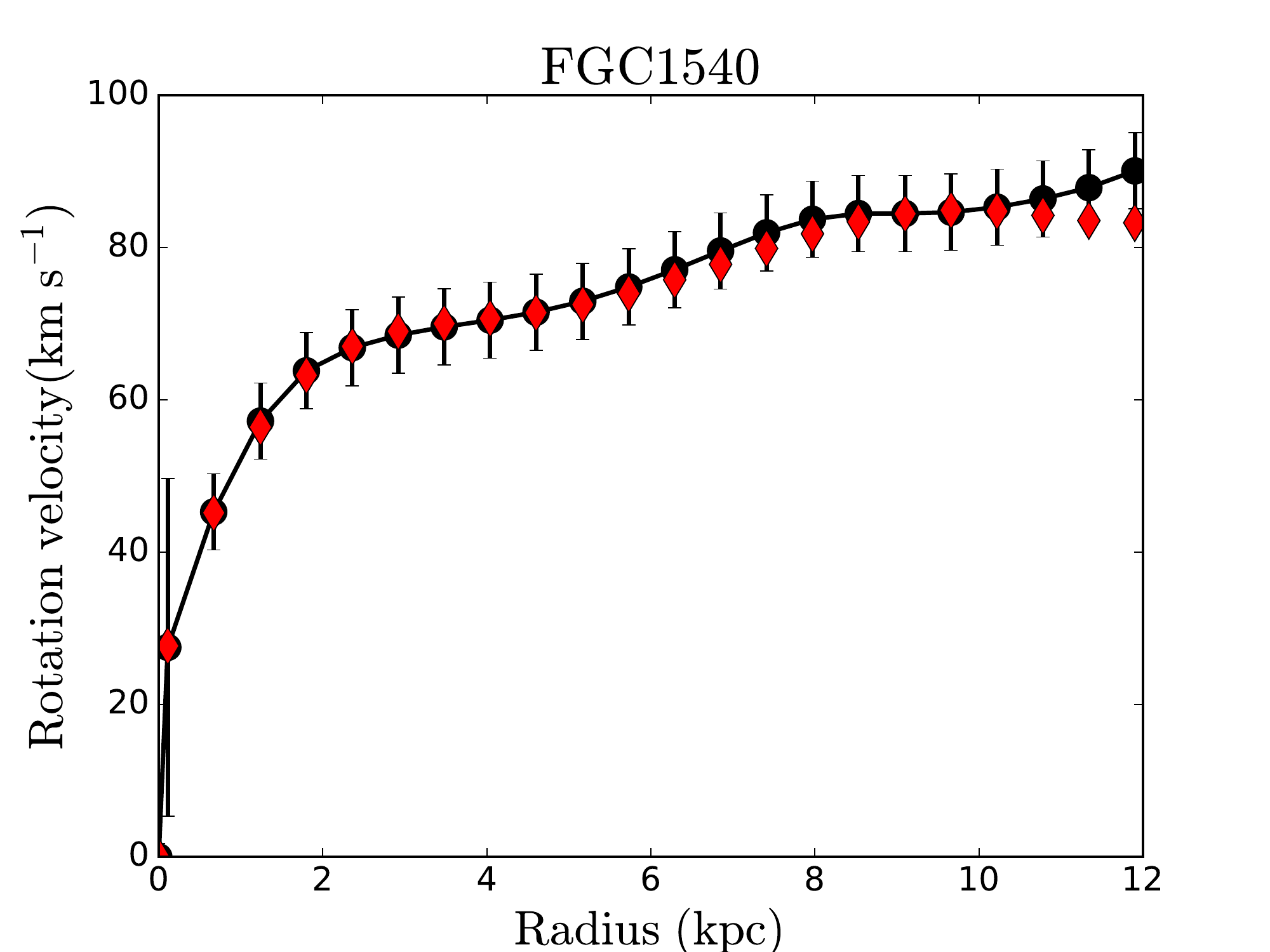}
\caption{ The derived rotation curve before (red diamonds) and after (black circles) the correction for pressure support.}
\label{fig:vcor}
\end{figure}

We fit a triple Gaussian profile to the radial surface density distribution and use a velocity dispersion, $\sigma \approx$ 10 km s$^{-1}$ (as derived from the FAT) and substitute them in Equation \ref{correction} to calculate the corrected rotation velocities. Fig \ref{fig:vcor} shows the rotation curves before (red diamonds) and after (black circles) the correction for pressure support.  In the inner regions, we find that the correction is small (less than 2.5 km s$^{-1}$) compared to the error bars on the rotation curve. Even in the outer region, the correction is only $\sim$ 1$\sigma$. We also note that the optical rotation curve matches broadly with the rotation curve derived from H{\sc i} distribution.

\section{Mass Models}
\label{models}

The net gravitational potential of the galaxy is equal to the superposition of the respective gravitational potentials due to the density distributions of the stars, gas and the dark matter halo. The density profiles of the stars and the gas can be modelled from observations; these, in conjunction with the observed rotation curve, may be used to constrain the density profile of the dark matter halo as discussed below.

\subsection{Stars}
\label{stell}

The rotational velocities due to the gravitational potentials of the thin and thick stellar discs (\S \ref{sdss}, \S \ref{spitzer}) were determined separately using the GIPSY task 
`ROTMOD' \citep{vanderhulst92}, and were added in quadrature in order to obtain the net contribution of the stellar disc, $V_{\rm \ast}$, to the total rotational velocity.
We determine the gravitational potential of the stellar disc for two different cases: SDSS $i$--band data and Spitzer IRAC 3.6 $\mu$m.


\subsection{Gas}

Similarly, the rotational velocity due to the gravitational potential of the gas, $V_{\rm gas}$, was obtained with `ROTMOD' using the de-projected H{\sc i} radial surface density profile as derived from FAT, assuming an exponential disc vertical structure. The H{\sc i} surface density values were scaled by a factor of 1.35 to account for the contribution of Helium. The contribution of molecular gas was neglected as the fraction of H$_{2}$ to H{\sc i} mass in late-type LSB galaxies is generally very small ($\sim $10$^{-3}$) \citep{matthews05}.

\subsection{Dark Matter Halo}

We model the dark matter halo first as (i) the observationally-motivated pseudo-isothermal (PIS) dark matter profile with a constant density core, 
and (ii) the cosmologically-motivated Navarro-Frenk-White (NFW) profile with a cusp central density \citep{navarro96} below.

The PIS dark matter density profile is given by

\begin{equation}
\rho_{iso} (R) =\rho_{0}[1+(R/R_{\rm C})^{2}]^{-1}
\end{equation}

where $\rho_0$ is the core density and $R_{\rm C}$  the core radius of the halo. The circular velocity due to the PIS halo is

\begin{equation}
V_{\rm iso} (R) = \sqrt{4 \pi \rho_{\rm 0} (R)  R_{\rm C}^{2}\Big [1-\dfrac{R_{\rm C}}{R} \tan^{-1}(\dfrac{R}{R_{\rm C}})\Big ]}
\end{equation}

\begin{figure*}
\centering
\subfloat{\includegraphics[width = 3in]{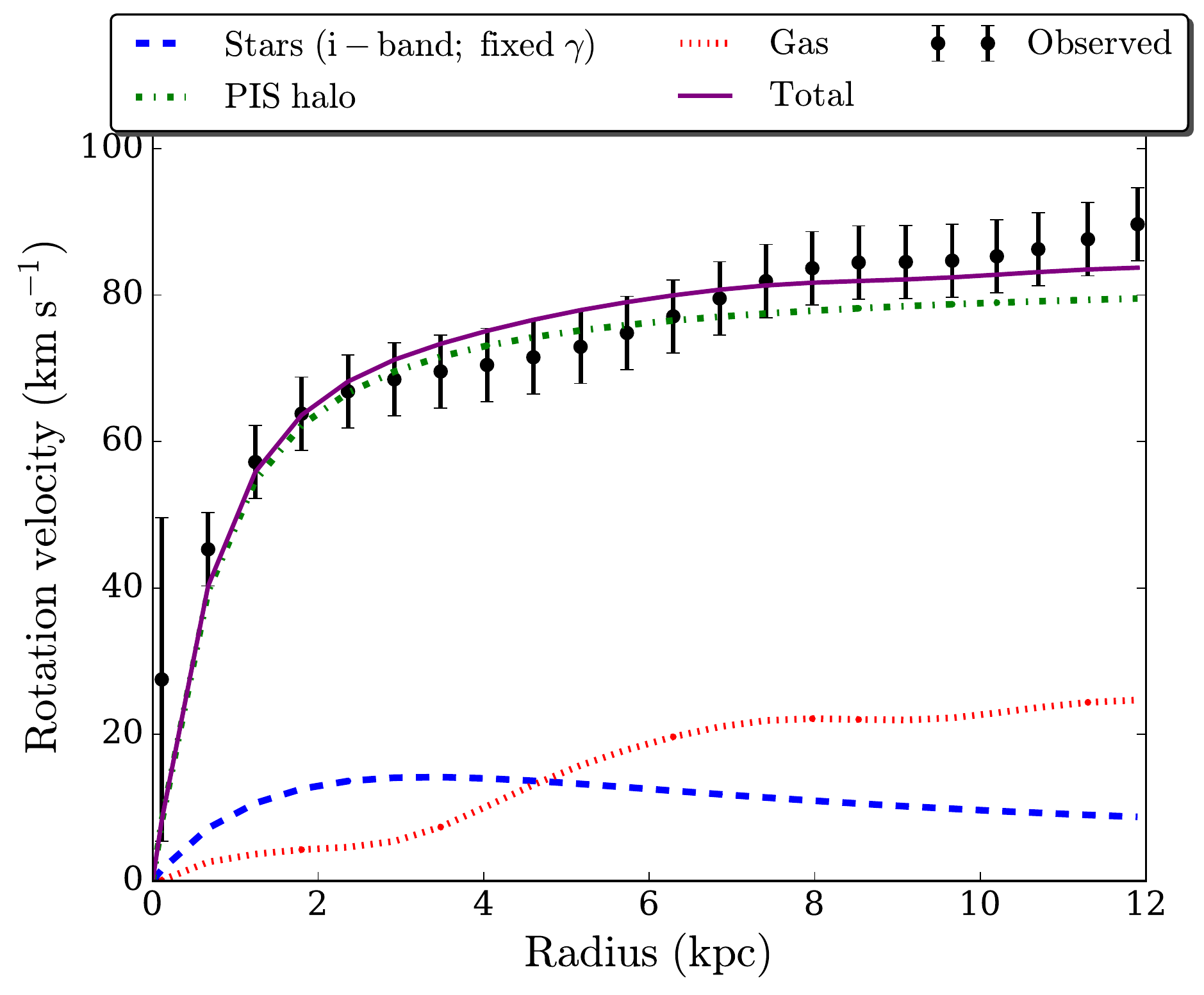}}
\subfloat{\includegraphics[width = 3in]{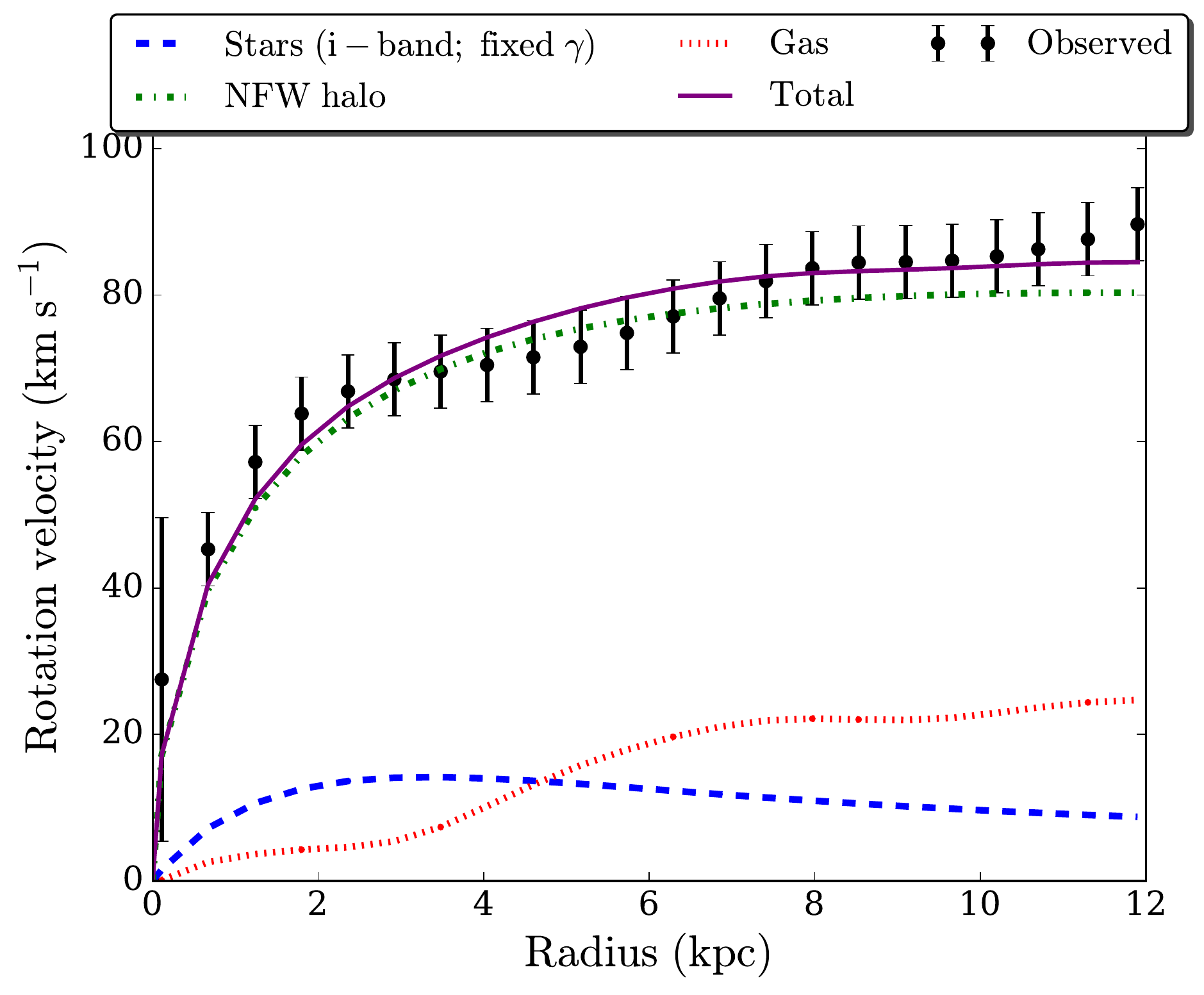}} \\
\subfloat{\includegraphics[width = 3in]{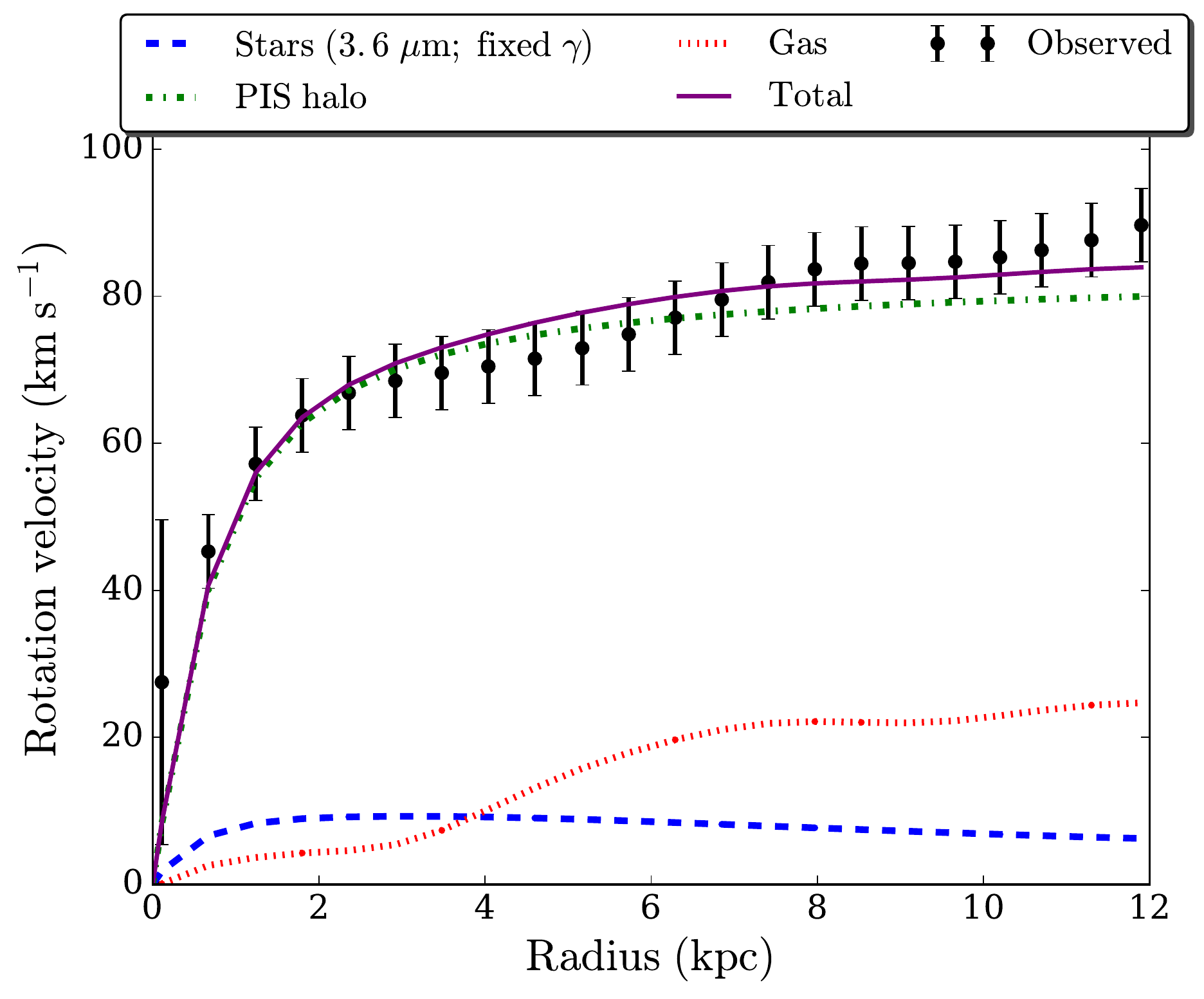}} 
\subfloat{\includegraphics[width = 3in]{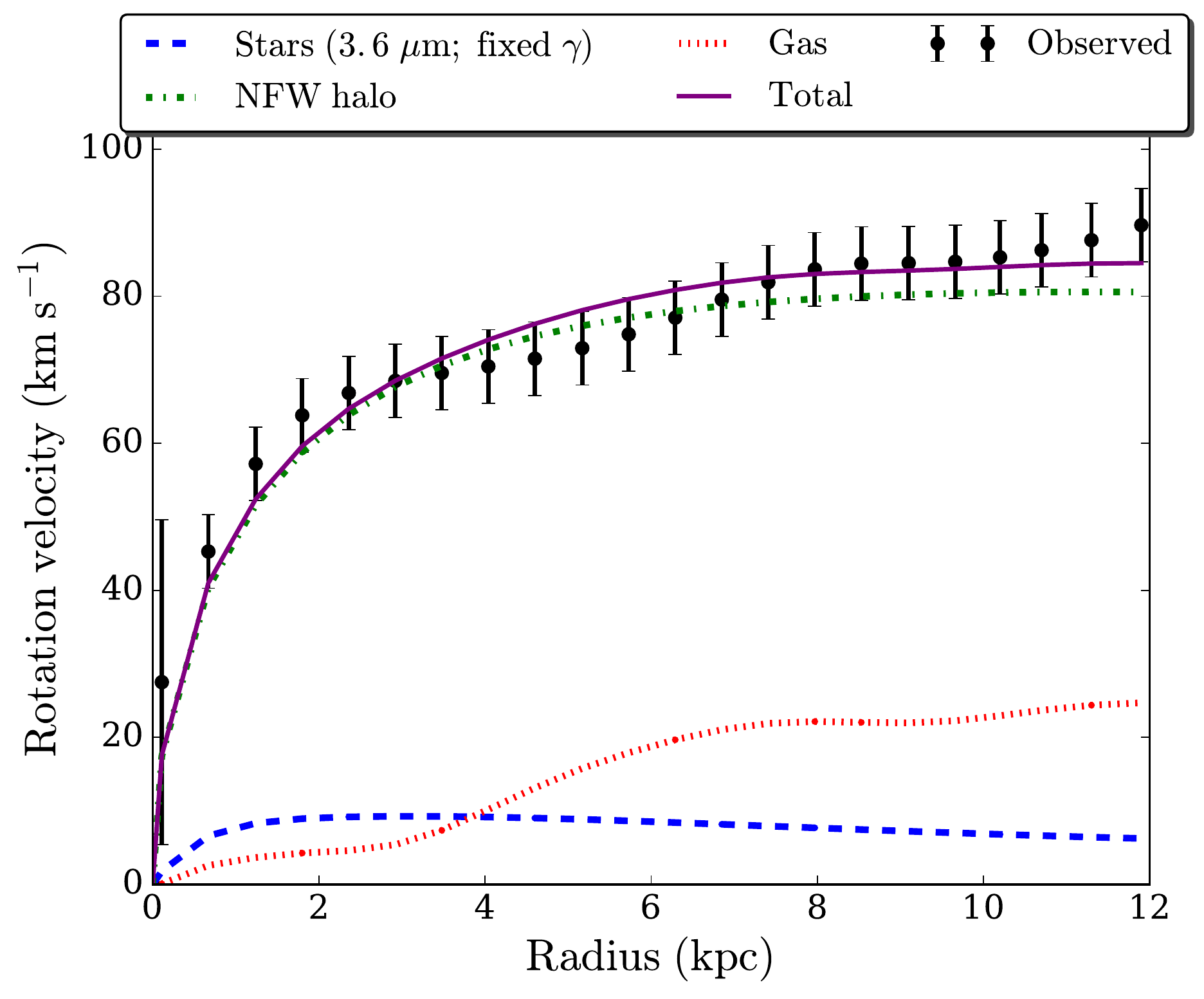}}
\caption{Upper panels: PIS and NFW halo based mass model for the FGC 1540 using the SDSS $i$--band data, Lower panels: PIS and NFW halo based mass model for the FGC 1540 using the Spitzer 3.6 $\mu$m data by fixing the $\gamma_{\ast}$ as derived from SPS models. The observed rotation curve is shown by black points with error bars. The blue dashed line shows the contribution of the stellar disk,  the  red dotted  line  shows  the  contribution  of  the  gas  disk and the dot-dashed line shows the contribution of the dark matter halo to the total rotation velocity. The solid line shows the quadrature sum of all of these components.}
\label{fig:constg}
\end{figure*}

The NFW halo has a density profile given by

\begin{equation}
\rho_{NFW} (R) =\dfrac{\rho_{\rm i}}{(R/R_{\rm s})(1+R/R_{\rm s})^{2}}
\end{equation} 

where $ R_{\rm s} $ is a characteristic radius and $ \rho_{\rm i} $ is related to the density of the universe at the time of collapse. The circular velocity due to the NFW halo is:

\begin{equation}
V_{NFW} (r) = V_{\rm 200} \sqrt{\dfrac{\ln (1+cx)-cx/(1+cx)}{x[\ln (1+c)-c/(1+c)]}}
\end{equation}

where $ c=R_{\rm 200}/R_{\rm s} $ and $ x=R/R_{\rm 200} $;
$ R_{\rm 200} $ being the radius at which the mean density of the halo is equal to 200 times the critical density and $ V_{\rm 200} $ the rotational velocity at $ R_{\rm 200} $.

Finally, mass models for FGC1540 were constructed using the observed rotation curve corrected for pressure 
support derived in the last section. Theoretically, the observed rotational velocity ($V_{\rm obs}$) can be shown to be equal to the quadrature sum of rotational velocities due to the 
respective gravitational potentials of the stars ($V_{\rm *}$), gas ($V_{\rm gas}$) and dark matter halo ($V_{\rm h}$):
$V_{\rm obs}^{2} = \gamma V_{\rm *}^{2} + V_{\rm gas}^{2} + V_{\rm h}^{2} $ \\
As discussed above, $V_{\rm *}$ and $V_{\rm gas}$ can be determined from the stellar (\S \ref{photometry}) and gas profiles (\S \ref{kinematics}) using the `ROTMOD' task in GIPSY. 
$V_{\rm h}$ is however not known, and is parametrized by 2 free parameters: {$\rho_{0}$, $R_{\rm C} $} for the PIS halo, {$c$, $x$} for the NFW halo. 
The least square fit to the observed rotation curve using the task 
`ROTMAS' in GIPSY \citep{vanderhulst92} was done in order to determine the best-fitting parameters of the dark matter halo.

\begin{figure*}
\noindent
\subfloat{\includegraphics[width = 3in]{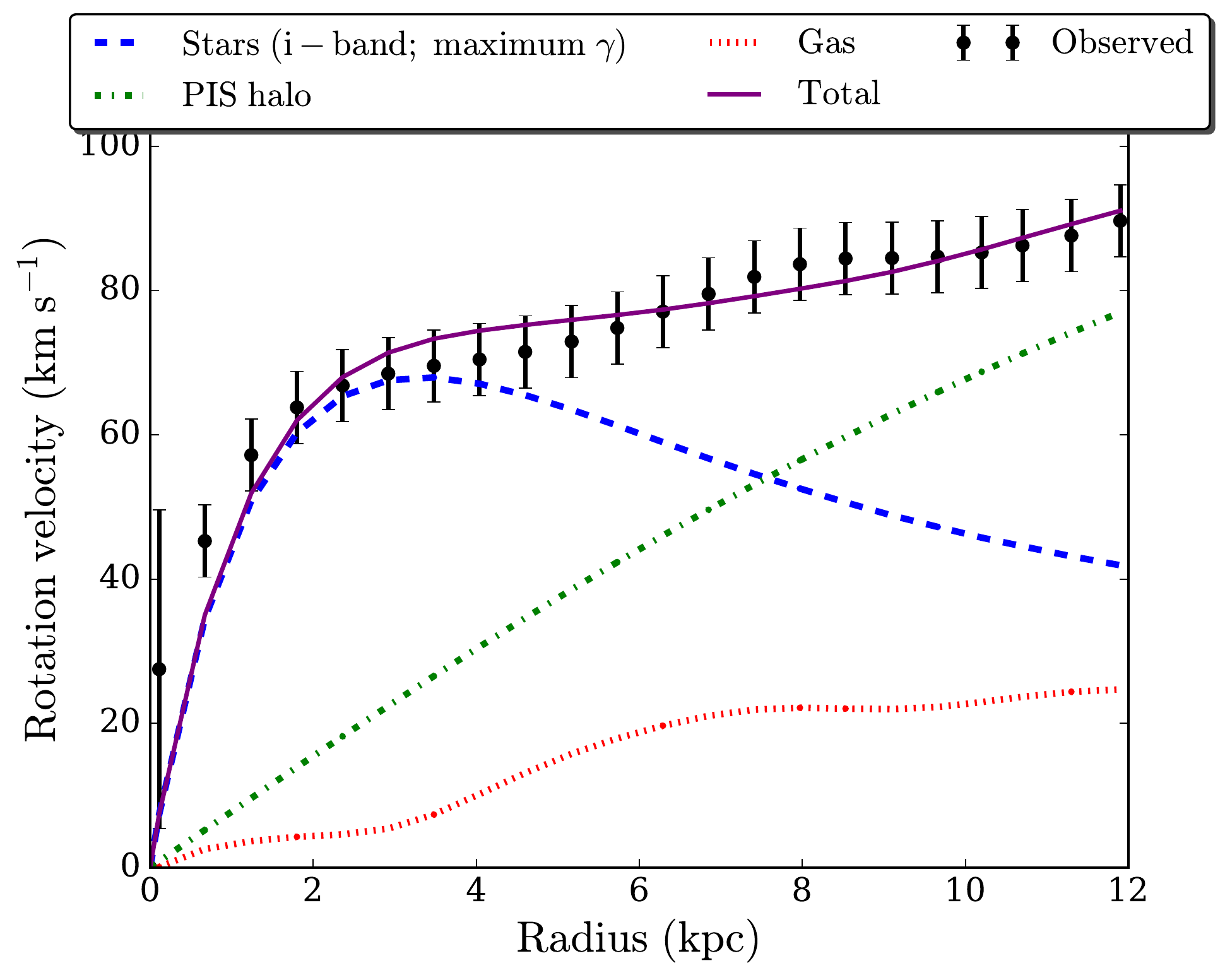}}
\subfloat{\includegraphics[width = 3in]{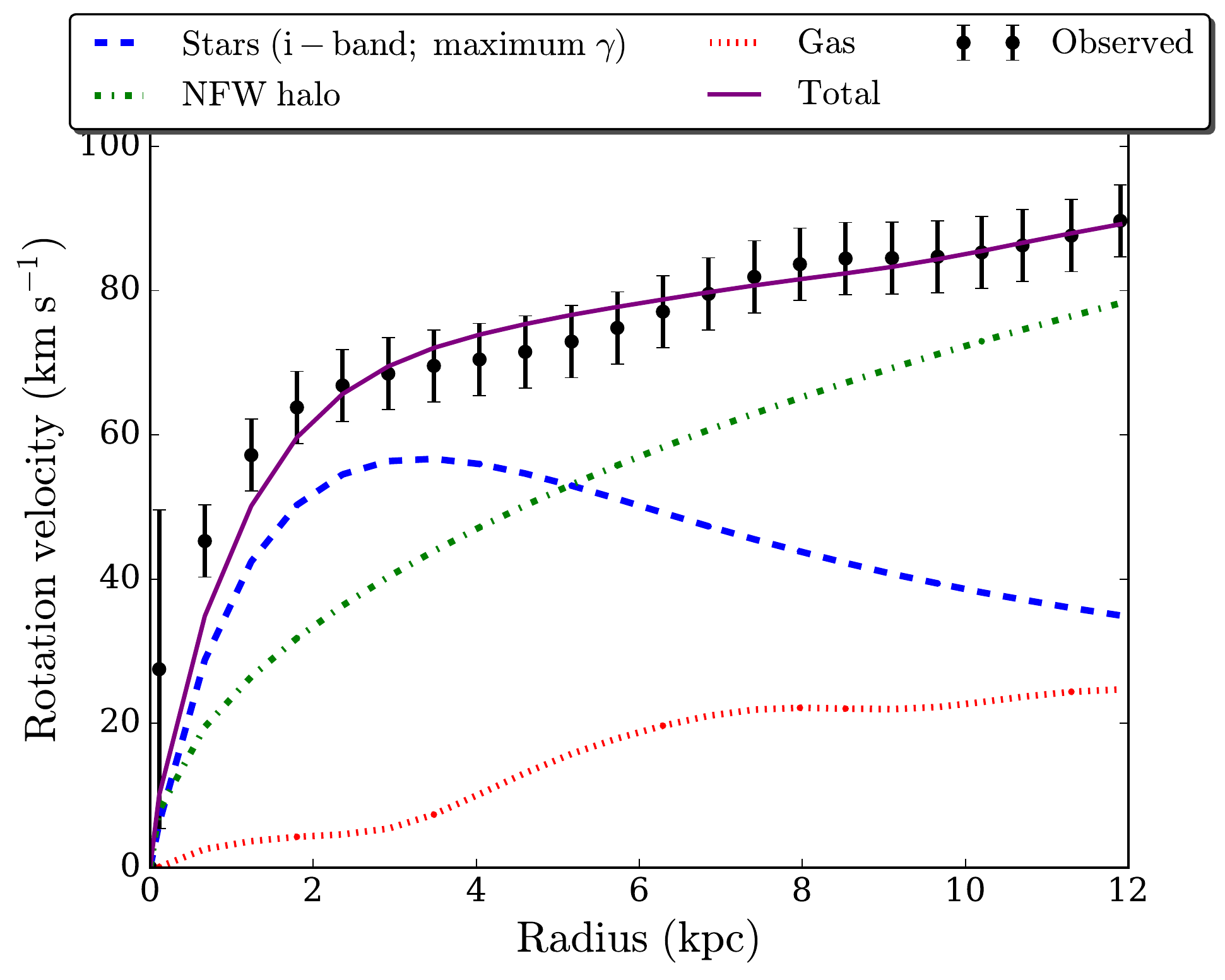}} \\
\subfloat{\includegraphics[width = 3in]{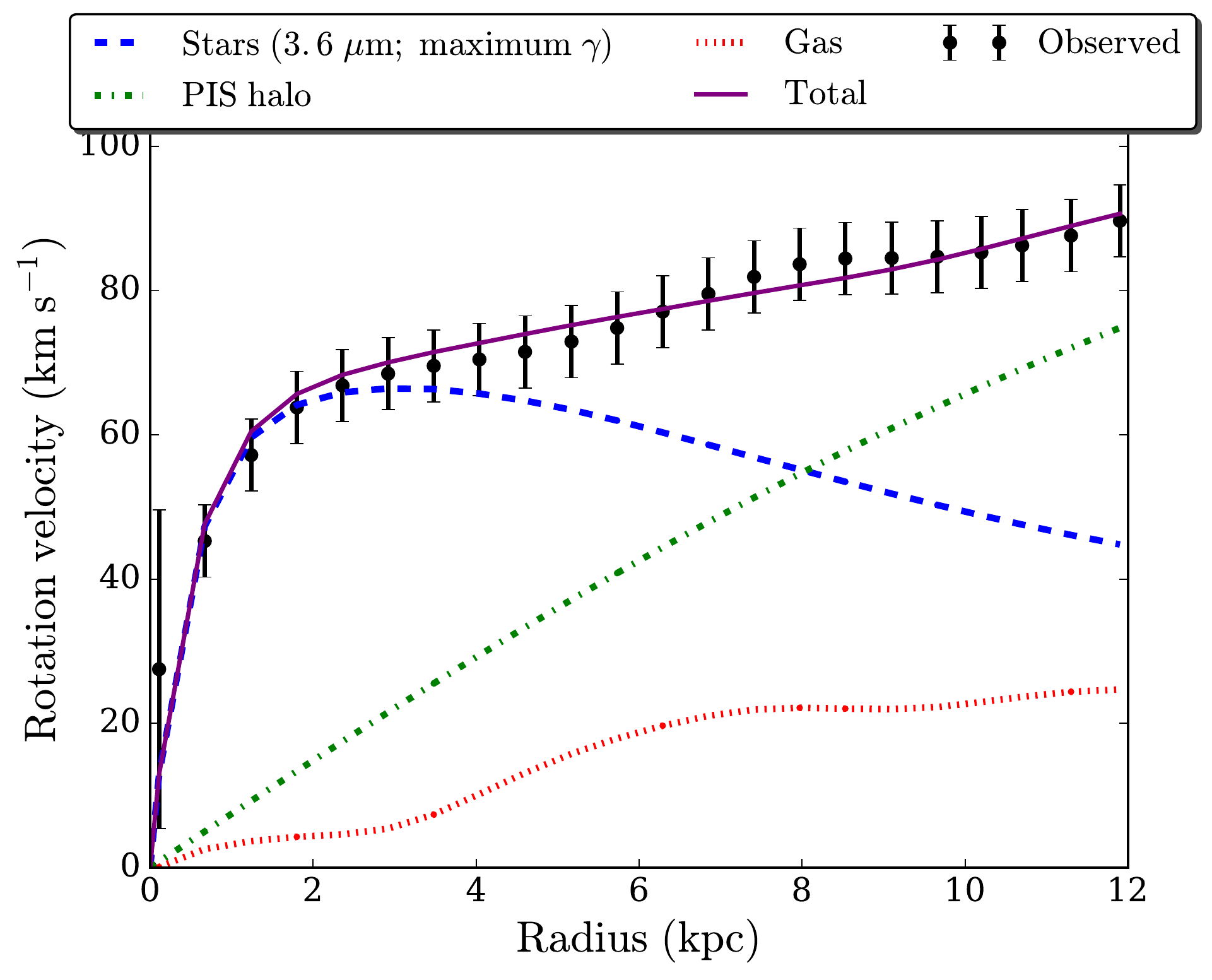}} 
\subfloat{\includegraphics[width = 3in]{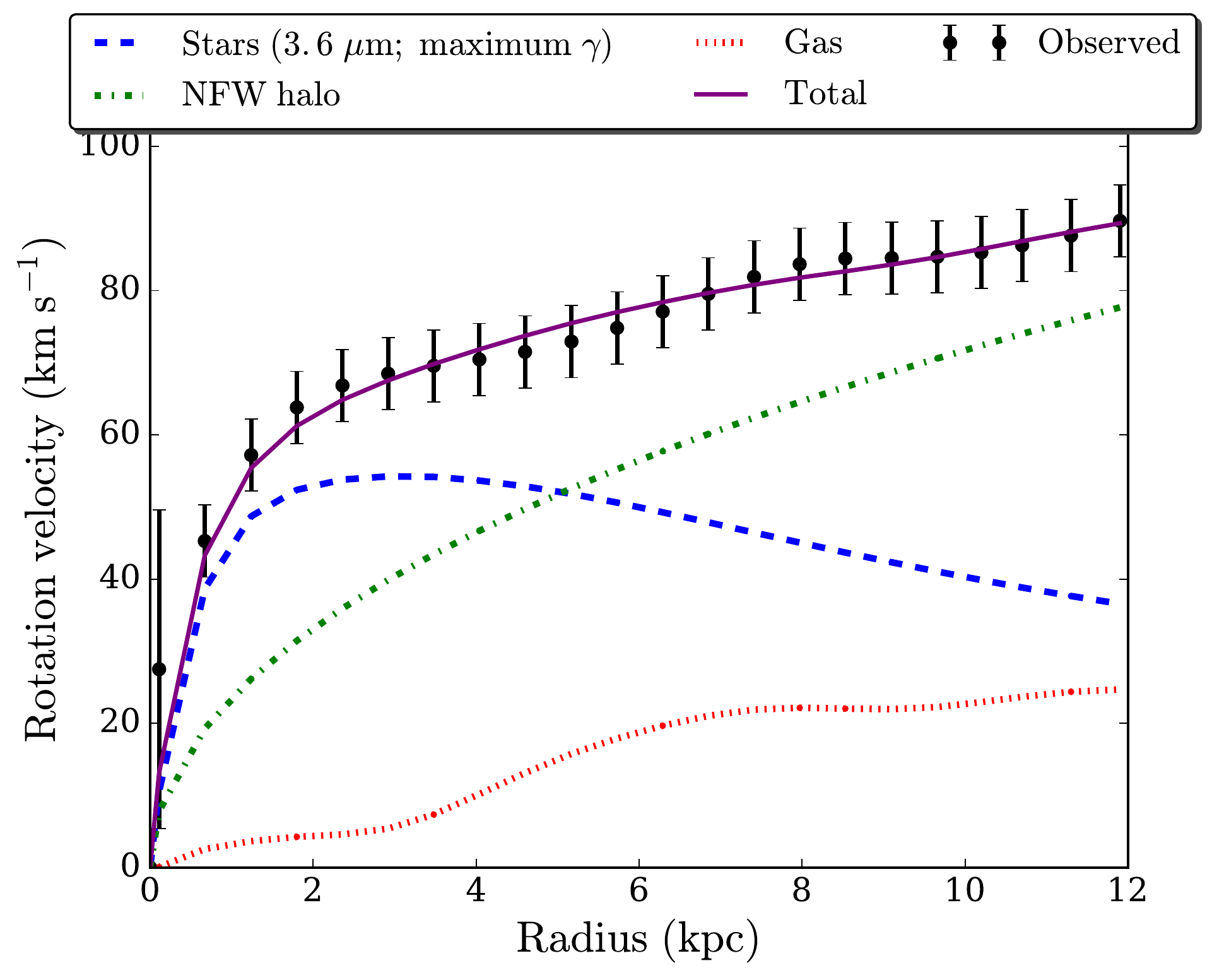}}
\caption{Upper panels: PIS and NFW halo based mass model for the FGC 1540 using the SDSS $i$--band data, Lower panels: PIS and NFW halo based mass model for the FGC 1540 using the Spitzer 3.6 $\mu$m data for the $maximum$ $disc$. The observed  rotation curve is shown by black points with error bars. The blue dashed line shows the contribution of the stellar disk,  the  red dotted  line  shows  the  contribution  of  the  gas  disk and the dot-dashed line shows the contribution of the dark matter halo to the total rotation velocity. The solid line shows the quadrature sum of all of these components.}
\label{fig:maxg}
\end{figure*}

 \begin{table*}
\small
\caption{Mass models with a PIS halo}
\label{table2}
\begin{tabular}{ p{3.5cm} p{2.0 cm} p{3.0cm} p{1.5 cm} p{1.5 cm} p{1.5 cm} p{1.5 cm} }
\\
\hline
 Model & Band & $\rho_{\circ}$ (10$^{-3}$ M$_{\odot}$ pc$^{-3}$) & $R_{\rm C}$ (kpc) & $R_{\rm C}$/ $R_{\rm D}$ & $\gamma_{\ast}$ & $\chi^{2}_{\rm red}$ \\ 
\hline
Constant $\gamma_{\ast}$ & Spitzer 3.6 $\mu$m & 319 $\pm$ 59  &0.63 $\pm$ 0.06 & 0.34 $\pm$ 0.10 & 0.35 & 0.47 \\
 & SDSS $i$--band & 308 $\pm$ 60  &0.64 $\pm$ 0.07 & 0.49 $\pm$ 0.11 & 0.54 & 0.51 \\ 
 \\
 Minimum disc &  & 337 $\pm$ 61  &0.61 $\pm$ 0.06 & 0.47 $\pm$ 0.10 & 0 & 0.45 \\
Minimum disc with gas &  & 280 $\pm$ 55  &0.70 $\pm$ 0.07 & 0.54 $\pm$ 0.10 & 0 & 0.64 \\
\\
 Maximum disc & Spitzer 3.6 $\mu$m & 3.1 $\pm$ 0.3  &13.9 $\pm$ 3.1 & 7.51$\pm$ 0.22 & 18.0 & 0.17 \\
 & SDSS $i$--band & 3.3 $\pm$ 0.5  &13.3 $\pm$ 4.7 & 10.3$\pm$ 0.35 & 12.4 & 0.52 \\  

  \\

\hline 
\end{tabular}

\end{table*}

\begin{figure*}
\noindent
\subfloat{\includegraphics[width = 3in]{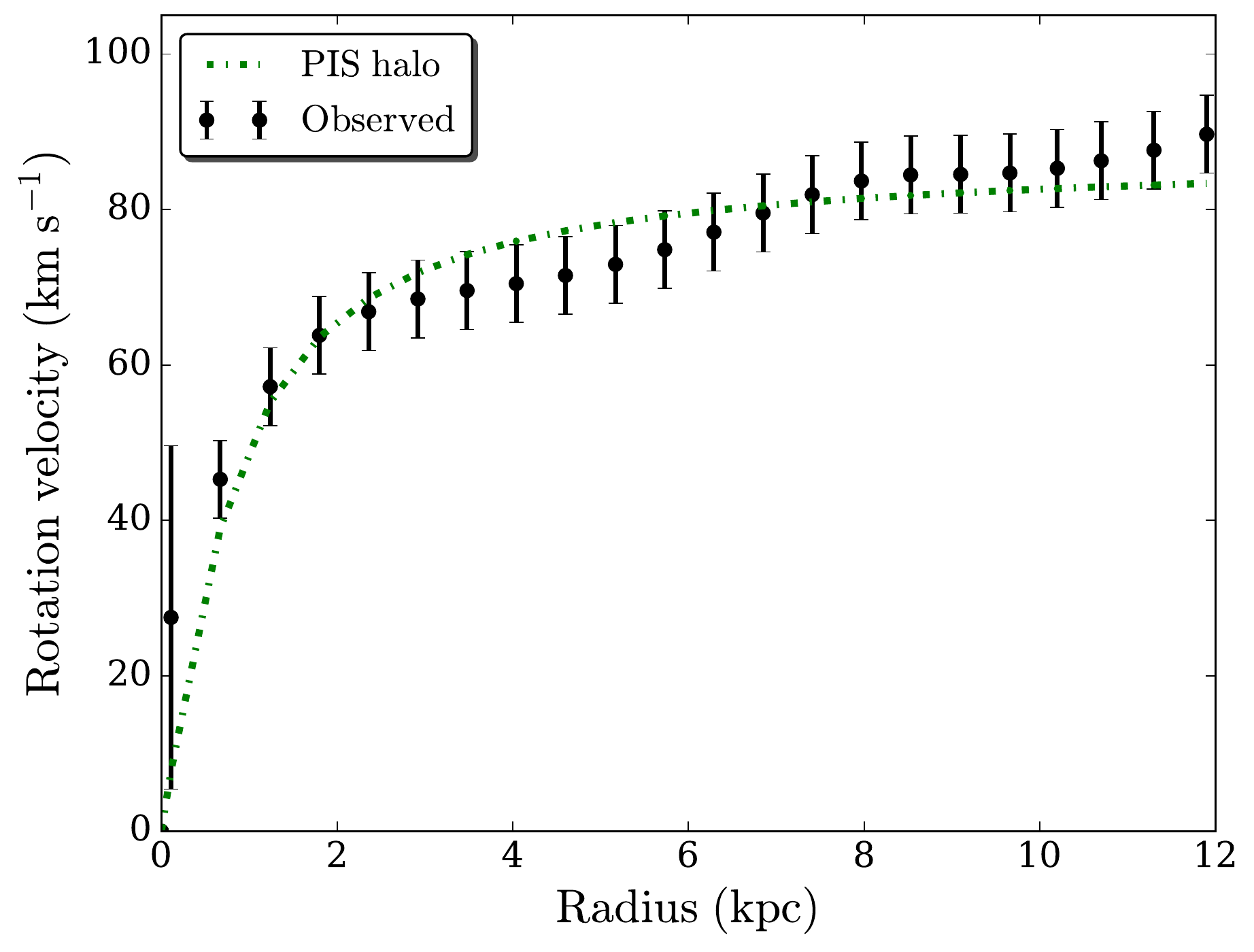}}
\subfloat{\includegraphics[width = 3in]{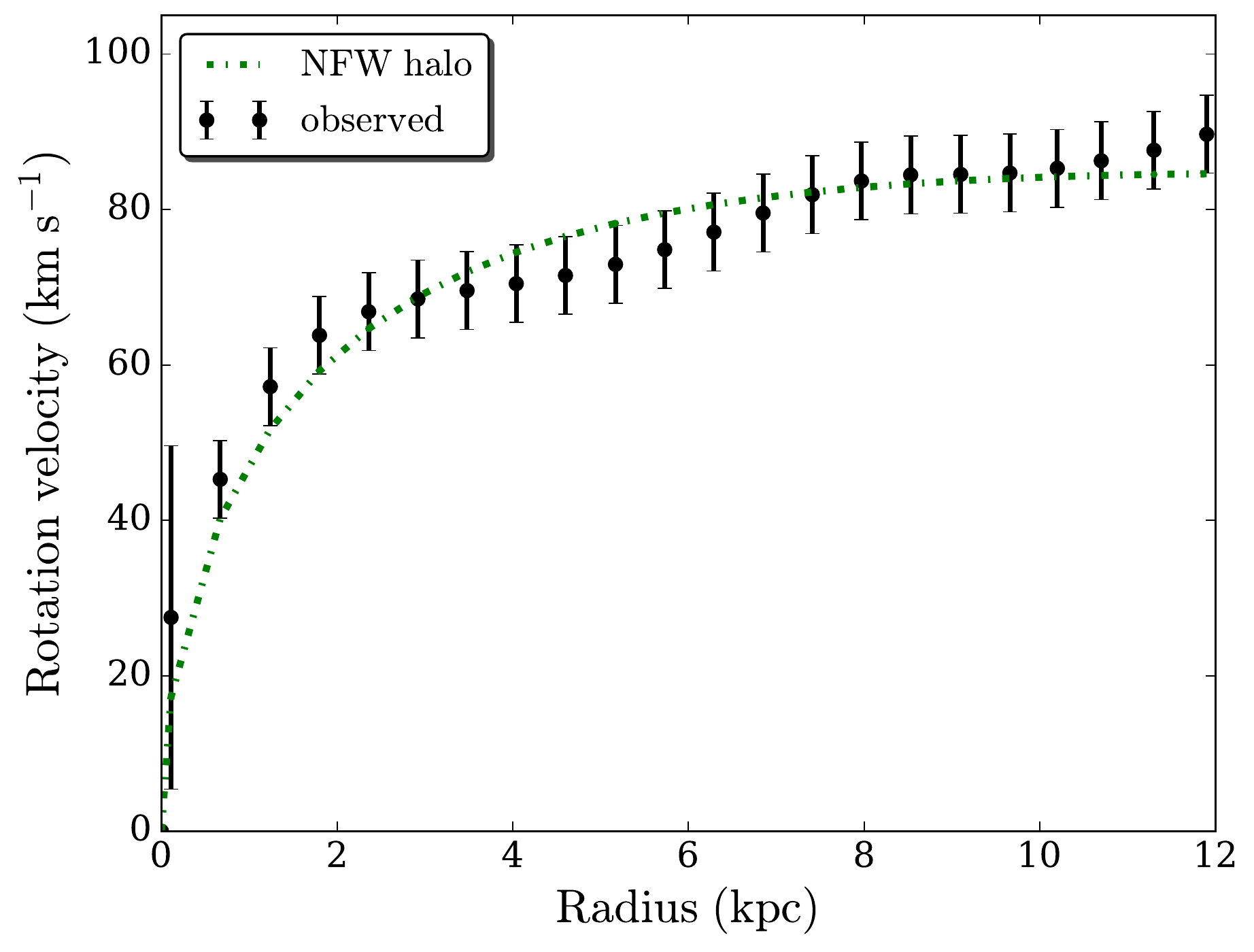}} \\
\subfloat{\includegraphics[width = 3in]{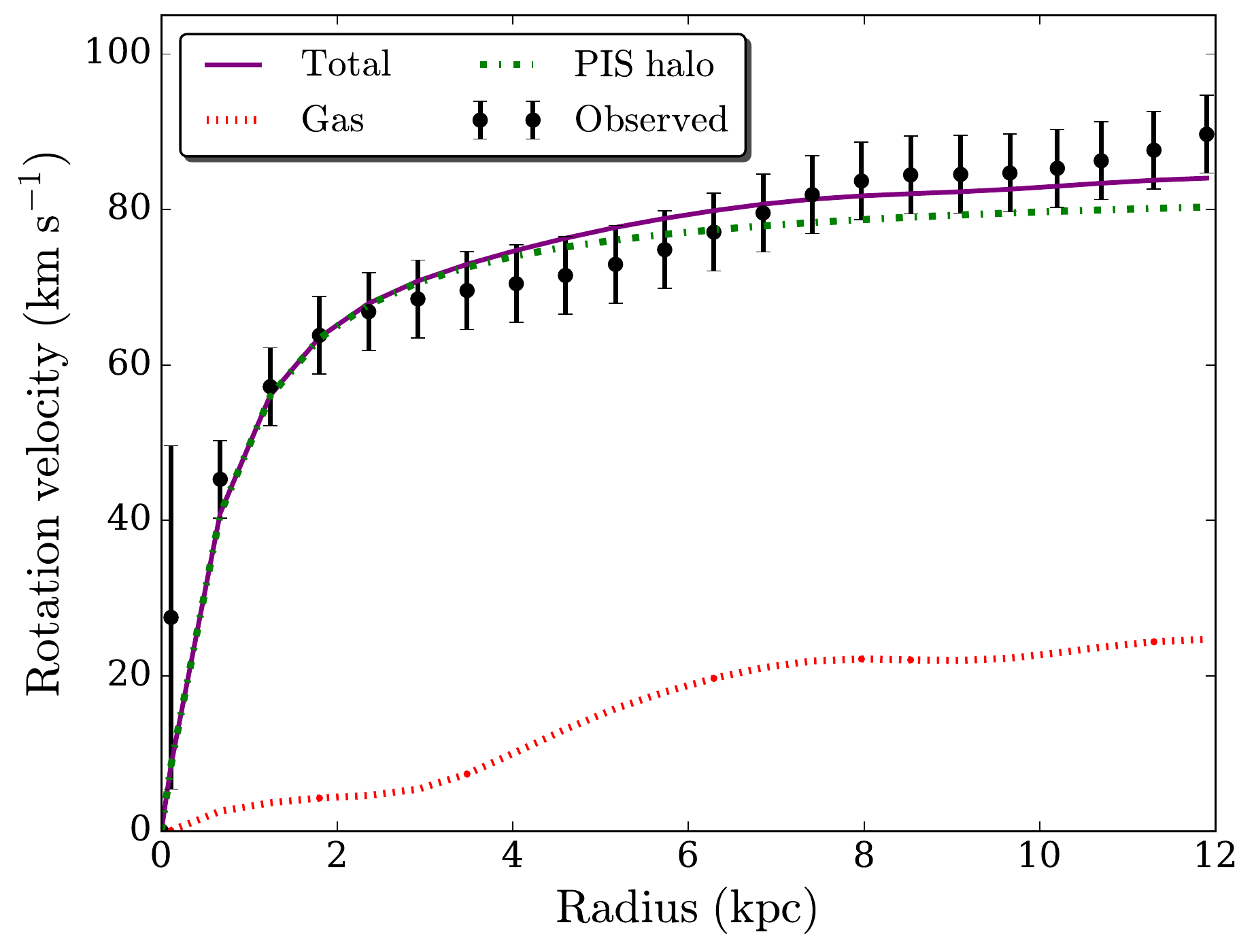}} 
\subfloat{\includegraphics[width = 3in]{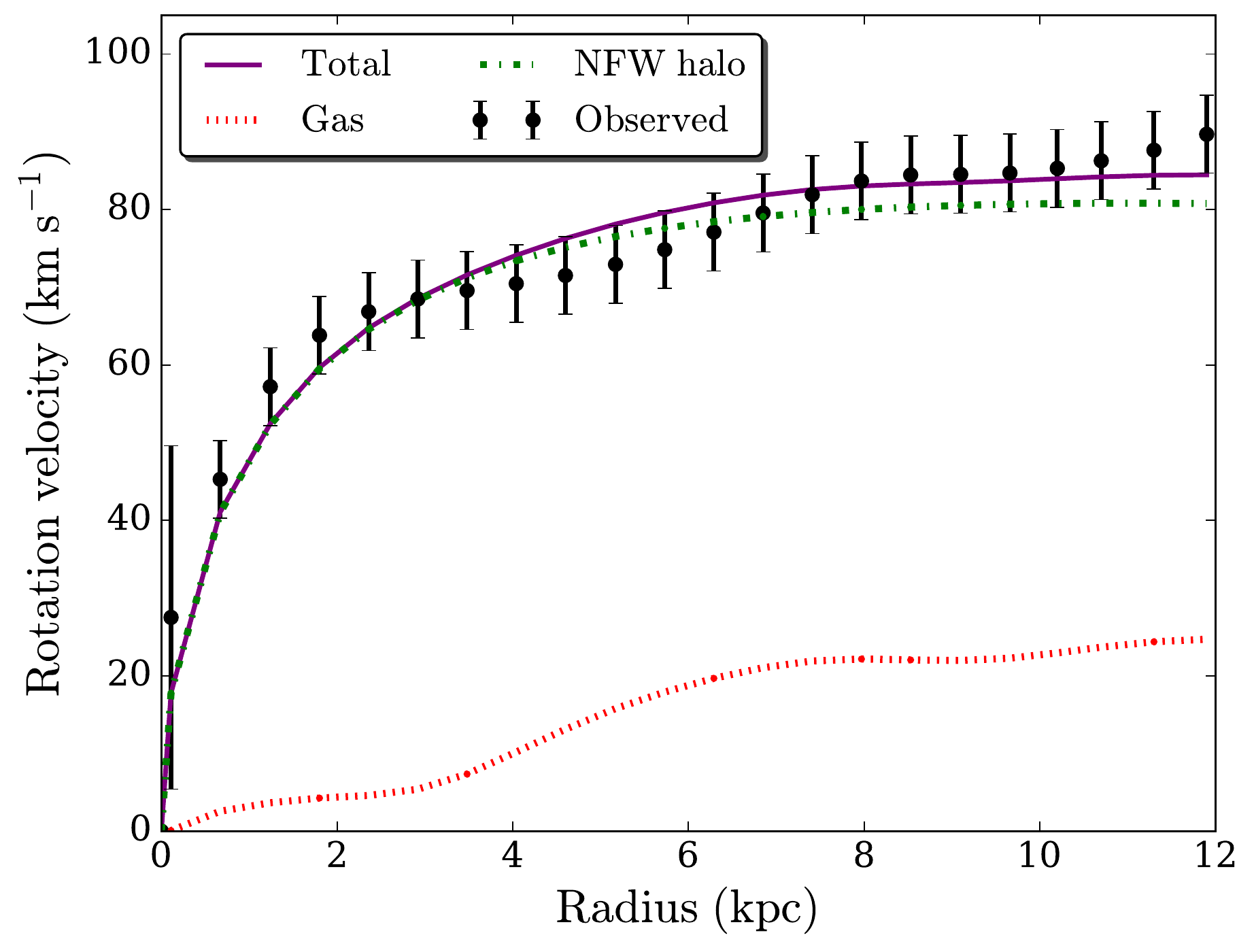}}
\caption{Upper panels: PIS and the NFW halo based mass model for the FGC 1540 for the $minimum$ $disc$, Lower panels: PIS and NFW halo based mass model for the FGC 1540 for the $minimum$ $disc$+ $gas$. The observed rotation curve is shown by black points with error bars. The  red dotted  line  shows  the  contribution  of  the  gas  disk and the dot-dashed line shows the contribution of the dark matter halo to the total rotation velocity. The solid line shows the quadrature sum of all of these components.}
\label{fig:mindisc}
\end{figure*}

\section{Results $\&$ Discussion}
\label{results}

The mass-to-light ratio ($\gamma_{\ast}$) is one of the major uncertainties while constructing mass models. 
We, therefore, construct mass models for the four different cases with (1) a fixed mass-to-light ratio $\gamma_{\ast}$ as dictated by the stellar population synthesis (SPS) models (2) a maximum disc (3) a minimum disc and (4) a minimum disc with gas. The maximum and minimum disc models respectively constrain the value of the dark matter density to be minimum and maximum at all $R$. We also considered a model allowing the mass-to-light ratio to be a free parameter. However, this did not yield any physically meaningful results and will therefore not be discussed in the following sections.

\subsection{Mass-to-light ratios and weighting}

\subsubsection{Constant $\gamma_{\ast}$}

In this case, we set the $\gamma_{\ast}$ equal to the value derived from the photometry and SPS models.

Figure \ref{fig:constg} [top panels] show the rotation curve decompositions for the best-fitting mass model using the SDSS $i$--band  stellar surface density profiles by fixing the $\gamma_{\ast}^{r}$ = 0.54 (\S \ref{sdss}).  For the PIS halo, the best-fit gives a central density $\rho_{\circ}$ = 0.308 $\pm$ 0.060 M$_{\odot}$ pc$^{-3}$ and a core radius $R_{\rm C}$ = 0.64 $\pm$ 0.07 kpc. The best-fitting model with an NFW halo corresponds to a concentration parameter $c$ = 9.1 $\pm$ 0.7 and characteristic radius $R_{\rm 200}$ = 49.5 $\pm$ 1.4 kpc. 

Figure \ref{fig:constg} [bottom panels] show the best-fitting mass model using the 3.6 $\mu$m stellar surface density profiles by fixing the $\gamma_{\ast}^{3.6}$ to 
0.35 (\S \ref{spitzer}). For the PIS halo, the best-fit corresponds to a central density $\rho_{\circ}$ = 0.262 $\pm$ 0.053 M$_{\odot}$ pc$^{-3}$ and a core radius 
$R_{\rm C}$ = 0.69 $\pm$ 0.07 kpc. The NFW halo now gives a concentration parameter $c$ = 8.6 $\pm$ 0.6 and  characteristic radius $R_{\rm 200}$ = 50.1 $\pm$ 1.5 kpc.

We note that the best-fitting values of the dark matter halo parameters
remain almost the same irrespective of the tracer of the stellar density profile used. This could be explained by the fact that the difference between the rotational velocity curves for the stellar potentials obtained from the respective bands is smaller than the error bars on the observed rotation curve.

\begin{table*}
\small
\caption{Mass models with NFW halo}
\label{table1}
\begin{tabular}{ p{3.5cm} p{2.0 cm} p{2.0cm} p{2.0 cm} p{1.5 cm} p{1.5 cm} p{1.5 cm} }
\\
\hline
 Model & Band & $c$ & $R_{\rm 200}$ (kpc) & $c_{\rm exp}$ & $\gamma_{\ast}$ & $\chi^{2}_{\rm red}$ \\ 
\hline
Constant $\gamma_{\ast}$ & Spitzer 3.6 $\mu$m & 9.4 $\pm$ 0.6  &49.4 $\pm$ 1.4 & 15 & 0.35 & 0.44 \\
 & SDSS $i$--band & 9.1 $\pm$ 0.7  &49.5 $\pm$ 1.4 & 15 & 0.54 & 0.47 \\ 
 \\
 Minimum disc &  & 9.6 $\pm$ 0.7  & 49.3 $\pm$ 1.3 & 15 & 0 & 0.45 \\
Minimum disc with gas &  & 8.8 $\pm$ 0.6  &52.6 $\pm$ 1.5 & 15 & 0 & 0.48 \\
\\
 Maximum disc & Spitzer 3.6 $\mu$m & 0.82 $\pm$ 0.46  &131.38 $\pm$ 38.87 & 15 & 12.0 & 0.11 \\
 & SDSS $i$--band & 0.87 $\pm$ 0.93  &128.5 $\pm$ 73.7 &15 & 8.64 & 0.52 \\  

  \\

\hline 
\multicolumn{7}{l}{Notes: $c_{\rm exp}$ is the expected concentration parameter from the scaling relations of $\Lambda$CDM cosmology. }

\end{tabular}

\end{table*}

\subsubsection{Maximum disc}

In this model, we scale the rotation curve due to the stellar component to a maximum value such that the
dark matter density is non-negative at all radii. As discussed, this sets a lower limit on the dark matter density and hence the contribution of the dark matter to the net gravitational potential of the galaxy.

Figure \ref{fig:maxg} [top panels] show the rotation curve decompositions with the mass model constructed using the SDSS $i$--band stellar surface density profiles by 
fixing the $\gamma_{\ast}^{i}$ to a maximum possible value. For the PIS halo, the best-fitting model gives a central density $\rho_{\circ}$ = 0.0033 $\pm$ 0.0005 M$_{\odot}$ pc$^{-3}$, the core radius $R_{\rm C}$ = 13.3 $\pm$ 4.7 kpc and mass-to-light ratio $\gamma_{\ast}$ = 12.4, whereas the NFW halo gives the concentration parameter $c$ = 0.87 $\pm$ 0.93, characteristic radius $R_{\rm 200}$ = 128.5 $\pm$ 73.7 kpc and mass-to-light ratio $\gamma_{\ast}$ = 8.64.

Figure \ref{fig:maxg} [bottom panels] show the best-fit mass model using the 3.6 $\mu$m stellar surface density profiles with the $Maximum$ $disc$ model. For the PIS halo, the best-fit gives a central density $\rho_{\circ}$ = 0.0035 $\pm$ 0.0003 M$_{\odot}$ pc$^{-3}$, the core radius $R_{\rm C}$ = 11.3 $\pm$ 1.6 kpc and $\gamma_{\ast}$ = 4.0. The NFW halo gives a reduced $\chi^{2}$ = 0.11, the concentration parameter $c$ = 0.20 $\pm$ 0.8,  characteristic radius $R_{\rm 200}$ = 213.9 $\pm$ 196.5 kpc and $\gamma_{\ast}$ = 4. 

We note that the values of the mass-to-light ratio $\gamma_{\ast}$ required for the best-fitting $Maximum$ $disc$ models are an order of magnitude higher than the corresponding values dictated by the stellar population syntheses models, which is not physical.

An alternative method of constructing $Maximum$ $disc$ models is to arbitrarily fix the $\gamma_{\ast}$ such that the observed rotational velocity $V_{\rm {rot}}$ at 2.2 $R_{\rm D}$  could be completely attributed to the baryonic disc without invoking the dark matter halo. However, the drawback with this model is one may end up with negative values of the dark matter densities as was found for the $Maximum$ $disc$ models with NFW halos for the superthin galaxies \citep{banerjee17}. Earlier work on LSBs including superthin galaxies using this alternative method indicated that LSBs mostly do not have a $Maximum$ $disc$ \citep{deBlok08} as is also the case for the superthins \citep{banerjee17}. 

\begin{figure}
  \centering
  \includegraphics[width=1.05\linewidth]{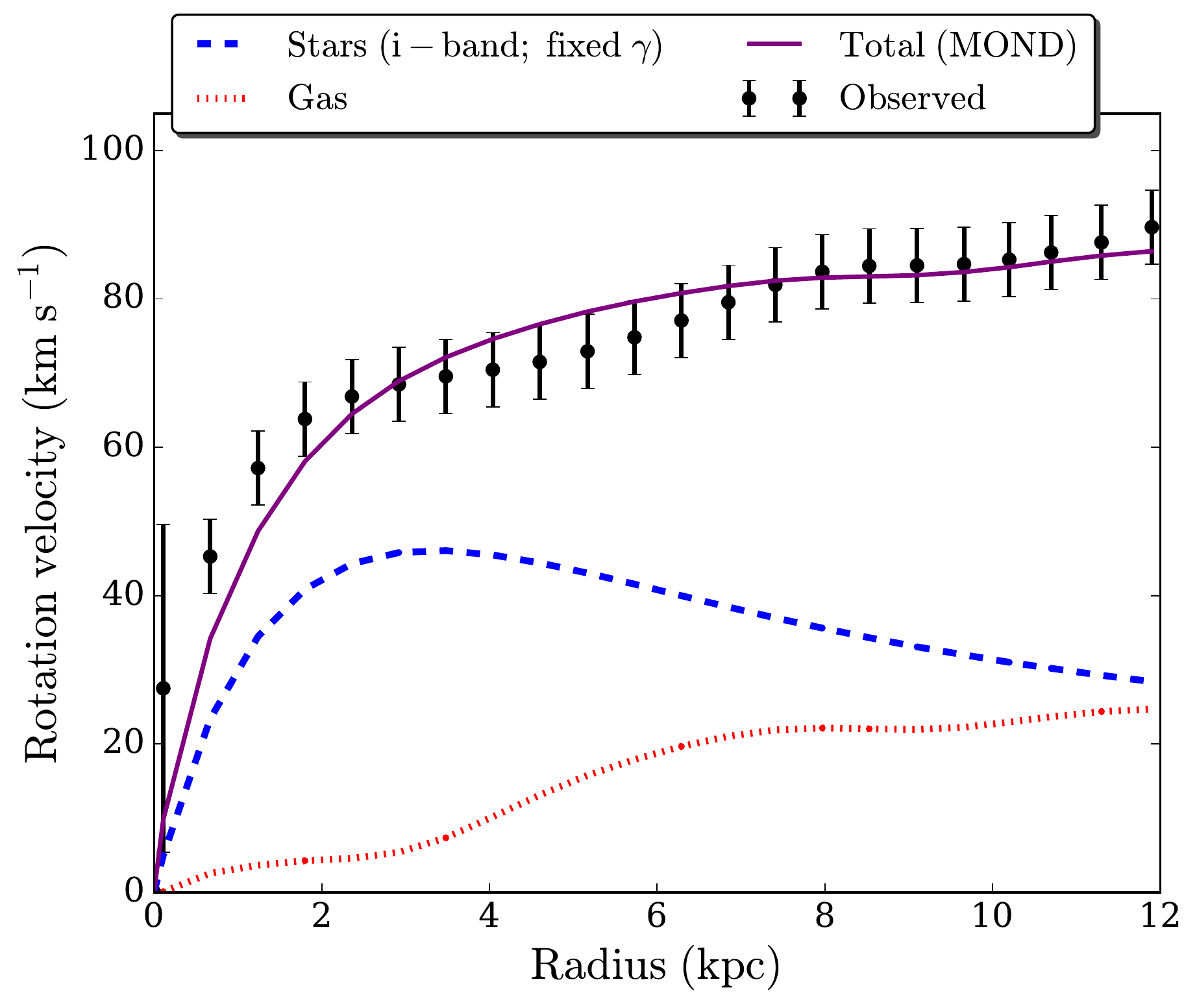}
  \includegraphics[width=1.05\linewidth]{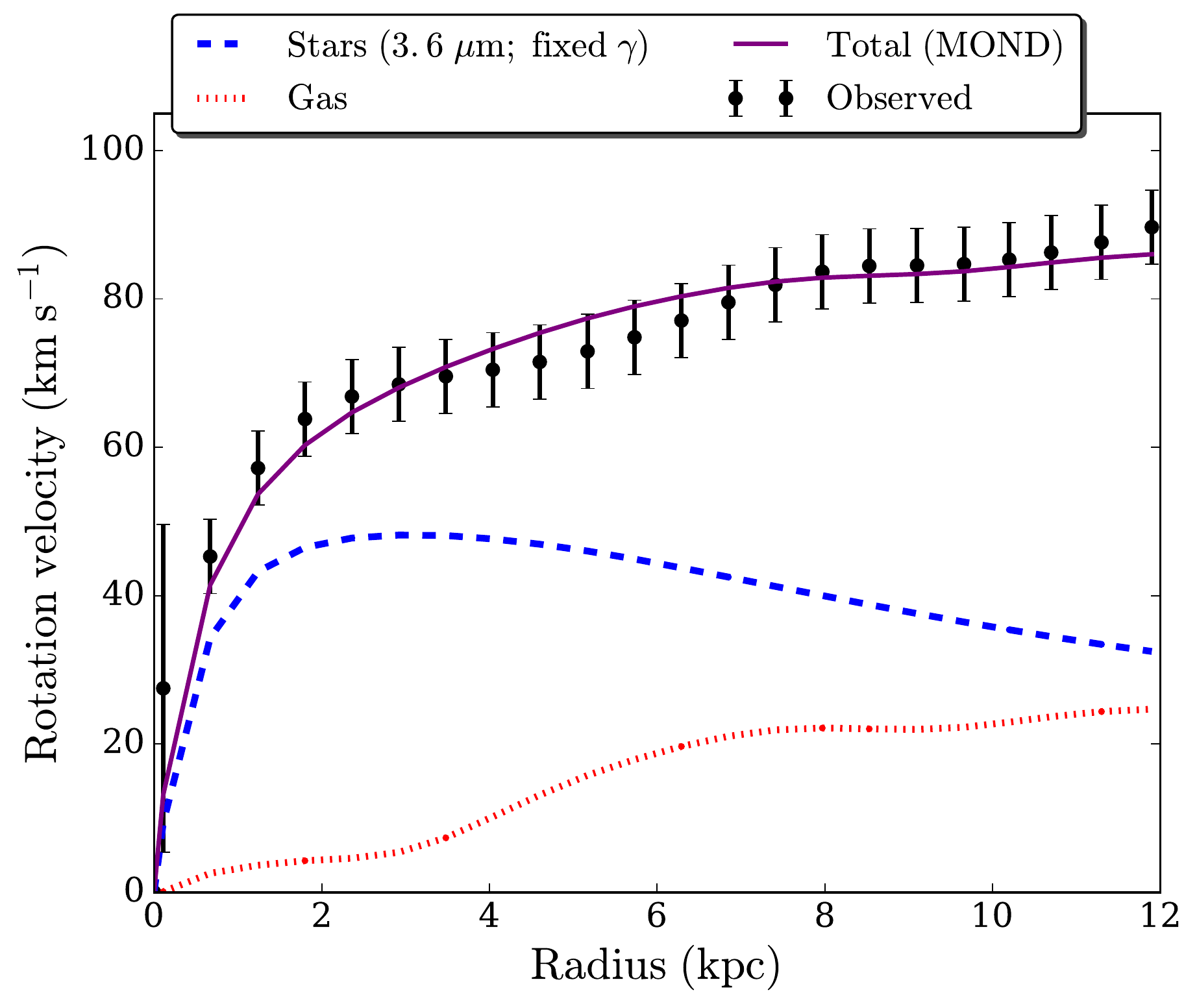}
  \caption{ Best fit for the Modified Newtonian Dynamics using the (a) SDSS $i$--band data and (b) the Spitzer 3.6 $\mu$m data.}
  \label{fig:mond}
\end{figure}

 \subsubsection{Minimum disc}
This model assumes that the contribution of stellar and gas disc to the rotation curve is zero and that the rotation curve is entirely due to the dark matter, thus giving an upper limit to the dark matter density. Figure \ref{fig:mindisc} [top panels] show the best-fit to the rotation curve for a PIS halo and and an NFW halo with the $Minimum$ $disc$ model. In this case, the PIS halo gives a central density $\rho_{\circ}$ = 0.280 $\pm$ 0.054 M$_{\odot}$ pc$^{-3}$ , 
and a core radius $R_{\rm C}$ = 0.70 $\pm$ 0.07 kpc. The NFW halo gives a concentration parameter $c$ = 8.8 $\pm$ 0.6 and  characteristic radius $R_{\rm 200}$ = 52.6 $\pm$ 1.5 kpc.

\subsubsection{Minimum disc with gas}

Here, we consider the contribution of gas (neutral hydrogen and helium) to the rotation curve is taken into account. But, we take the contribution of stellar disc to be zero ($\gamma_{\ast}$ = 0). Figure \ref{fig:mindisc} [bottom panels] show the mass models with zero stellar contribution. In this case, PIS halo gives a central density $\rho_{\circ}$ = 0.337 $\pm$ 0.061 M$_{\odot}$ pc$^{-3}$ and the core radius $R_{\rm C}$ = 0.61 $\pm$ 0.06 kpc. The NFW halo gives a concentration parameter $c$ = 9.6 $\pm$ 0.7 and  characteristic radius $R_{\rm 200}$ = 49.3 $\pm$ 1.3 kpc. 

We note, the best-fitting values of the dark matter halo parameters for both the $Minimum$ $disc$ models match within the error bars those from the model with fixed $\gamma_{\ast}$, implying that the contribution of the disc to the net gravitational potential is negligible. This is, again, in compliance with the trend found for other superthin galaxies \citep{banerjee17} and also LSBs in general \citep{deBlok08}.

 The results of the best-fitting models with PIS halo and NFW halo are summarized in Tables \ref{table2} and \ref{table1}.

\begin{table}
\small
\caption{Mass models with MOND}
\label{table_mond}
\begin{tabular}{ p{2.5 cm} p{1.5 cm} p{1.5 cm} p{1.5cm}  }
\\

\hline
 Band & a$_{0}$ & $\gamma_{\ast}$ & $\chi^{2}_{\rm red}$  \\ 
 \hline
 
 SDSS $i$--band & 3248 $\pm$ 708 & 5.70 $\pm$ 1.40 & 0.78  \\
 Spitzer 3.6 $\mu$m & 2693 $\pm$ 361 & 9.46 $\pm$ 1.30 & 0.31  \\

\hline

\multicolumn{4}{l}{Acceleration per unit mass (a$_{0}$) is in units of km$^{2}$ s$^{-2}$ kpc$^{-1}$. }
\end{tabular}

\end{table}

We obtain good fits to the observed rotation curve of the superthin galaxy FGC1540 for both the PIS and the NFW dark matter halo models using both the  SDSS $i$--band data and Spitzer 3.6 $\mu$m data, with the values for different parameters obtained using the SDSS $i$--band data and Spitzer 3.6 $\mu$m data for the stellar contribution matching within error-bars. The reduced $\chi^{2}$ is less than 1 for both the PIS and the NFW dark matter halo models, which indicates that error bars on rotation velocity as derived by the FAT package are likely to be overestimated.

We find that except for the $Maximum$ $disc$ models, the ratio of core radius of halo to the scale length of the optical disc ($R_{\rm C}$/ $R_{\rm D}$) is $\sim$ 0.5 for the PIS dark matter halo. This indicates that the core of the pseudo isothermal halo is compact, which is consistent with the compact cores ($R_{\rm C}$/ $R_{\rm D}$ < 2) obtained by \citet{banerjee17} for 3 superthin galaxies (UGC7321, IC5249 and IC2233), but is in contrast with the case of high surface brightness galaxies. \citet{banerjee13} showed that the compact nature of the dark matter halo is fundamentally responsible for the superthin vertical structure of the stellar disc in the prototypical superthin UGC7321. Their analytical calculations showed that for a dark matter halo with a given asymptotic rotational velocity or, in other words, a given total mass, the mean vertical scaleheight-to-radial scalelength ratio remains less than one ($z_{\rm 0}/R_{\rm D} < 1$) only for compact dark matter halos ($R_{\rm C}/R_{\rm D} < 2$). This is because for a dark matter halo with a compact core i.e., a core size comparable with the size of the stellar disc, the major mass fraction of the halo is concentrated within the galactic stellar disc, and hence the dark matter halo regulates the disc dynamics in the inner galaxy as well. In contrast, a non-compact dark matter halo dominates the galaxy dynamics mainly at the outer radii, and as a result, its vertical gravitational field cannot strongly act on the stellar disc and render it superthin. Hence, in the case of FGC1540 as well, the compact dark matter halo may be primarily responsible for regulating the superthin vertical structure of the stellar disc.


We can also compare the expected concentration parameter from the scaling relations as predicted from $ \Lambda $CDM cosmology.  The expected concentration parameter \citep[see e.g.][]{bullock01,bottema15} in terms of maximum rotation velocity is given by

\begin{equation}
c_{\rm exp} = 55.74 \times (V_{\rm max}(km\  s^{-1}))^{-0.2933}
\end{equation}

where $c_{\rm exp}$ is the expected concentration parameter and V$_{\rm max}$ the maximum rotation velocity. From the rotation curve, the maximum rotation velocity is 84 km s$ ^{-1} $, which gives expected concentration parameter to be 15, which differs from the concentration parameter that was obtained from mass modelling $\sim 9.0$. However, we could also derive the value of $c$ from the empirical relation between $c$--$V_{\rm 200}$ from the \citep{deBlok03, mcGaugh07} by substituting the $V_{\rm 200}$ with the $V_{\rm ISO}$($R_{\rm max}$) \citep[e.g.][]{oh15}.We obtain a concentration parameter value of $\sim$ 8.9, which matches with the value obtained from the mass modelling.



\subsection{Mass models with MOND formalism}

Modified Newtonian Dynamics (MOND) was proposed as an alternative hypothesis to explain the observed rotation curves of galaxies \citep{milgrom83}. This theory postulates that Newtonian dynamics breaks down at small accelerations. In contrast to Newtonian dynamics, MOND-ian dynamics rule out the presence of dark matter and takes into account the self-gravity of the stars and gas only in galactic dynamics studies. Here, the two free parameters are the stellar mass-to-luminosity ratio ($\gamma_{\ast}$) and the acceleration per unit length (a$_{0}$).

Fig. \ref{fig:mond} shows the best-fitting rotation curve for the MOND formalism using both the SDSS $i$--band data and the Spitzer 3.6 $\mu$m data. The parameters obtained for the MOND formalism are summarized in Table \ref{table_mond}.  In compliance with the earlier studies of superthin galaxies, the best-fitting $\gamma_{\ast}$ is not unusually high compared to those dictated by stellar population synthesis models. The best-fit acceleration parameters (a$_{0}$) do not differ significantly with the acceleration parameter (3734 km$^{2}$ s$^{-2}$ kpc$^{-1}$) expected from MOND theory. This possibly indicates that the superthins comply with mass models under the MOND formalism.

\section{Summary and Conclusions}
\label{summary}

We present the GMRT 21cm radio-synthesis observation of the superthin galaxy, FGC1540, and also its rotation curve and H{\sc i} surface density modelled using the publicly available software FAT (Fully Automated Tirrific). Using the SDSS $i$--band and Spitzer 3.6 $\mu$m data in conjunction with the H{\sc i} data, we also construct mass models for FGC1540.
We find that both the pseudo isothermal halo and NFW halo fit equally well for the dark matter; the best fitting models with the PIS halo profile indicates a compact core, where the best fit core radius ($R_{\rm C}$) is approximately half the stellar disc scale length ($R_{\rm D}$). This is in contrast with the case of high surface brightness galaxies where the halo core radius is typically 3-4 times that of the stellar scale length. \citet{banerjee17} also found that for 3 superthin galaxies (UGC7321, IC5249, and IC2233), the best fitting mass model with a PIS dark matter density profile indicate a compact dark matter halo (with $R_{\rm C}$/ $R_{\rm D}$ < 2 ). A compact dark matter halo strongly regulates the distribution of stars in the vertical direction as the vertical thickness of a galactic stellar disc is determined by the balance between the vertical self-gravitational force and pressure \citep[see e.g.][]{banerjee13, ghosh14, garg17}. This supports the suggestion that the compact core of the dark matter halo is responsible for the existence of superthin galaxies.
\section*{Acknowledgements}
This paper is based in part on observations taken with the GMRT. We thank the staff of the GMRT who made these observations possible. The GMRT is run by the National Centre for Radio Astrophysics of the Tata Institute of Fundamental Research. We acknowledge the usage of the HyperLeda database (http://leda.univ-lyon1.fr). The work of Russian researchers was supported by the Russian Science Foundation grant 14--12--00965.





\bibliographystyle{mnras}
\bibliography{superthin} 



\appendix


\bsp	
\label{lastpage}
\end{document}